\documentclass[a4paper,11pt]{article}
\usepackage{jheppub} % for details on the use of the package, please see the JINST-author-manual
\usepackage{lineno}
\usepackage{epsfig,xcolor,amsmath,tikz}
\usepackage{graphicx,subfig,subcaption}
\usepackage{epsfig}
\usepackage{color}
\usepackage{physics}
\newcommand{\x}{\vec{x}} 

\newcommand{\e}{\text{e}}
\newcommand{\colorgapunderline}[3][2.5pt]{%
  \tikz[baseline=(N.base)]{
    \node[inner sep=0pt, outer sep=0pt] (N) {$\displaystyle #3$};
    \draw[#2, line width=0.65pt] 
      ([yshift=-#1] N.south west) -- ([yshift=-#1] N.south east);
  }%
}

\arxivnumber{2605.09736} % if you have one
\title{Taming the infrared in de Sitter space: autonomous equations, stochastic approach, and Borel resummation} %Quantum field theory and stochastic approach in de Sitter spacetime: Autonomous equation and Borel resummation}}

\author[a,b]{Alexander Kamenshchik,}
\emailAdd{kamenshchik@bo.infn.it}
\affiliation[a]{Department of Physics and Astronomy ``A. Righi'', University of Bologna, \\ via Irnerio 46,  40126 Bologna, Italy}
\affiliation[b]{INFN, section of Bologna, \\ viale Berti Pichat 6/2, 40127 Bologna, Italy}
\author[c]{Polina Petriakova,}
\emailAdd{petriakova@theor.jinr.ru}
\affiliation[c]{Bogoliubov Laboratory of Theoretical Physics, Joint Institute for Nuclear Research, \\ Joliot-Curie Str. 6, 141980 Dubna, Russian Federation}
\author[d]{and Tereza Vardanyan}
\emailAdd{tereza.vardanyan@aanl.am}
\affiliation[d]{Alikhanyan National Laboratory (Yerevan Physics Institute), \\
Alikhanian Br. Str. 2, 0036 Yerevan, Armenia}
\abstract{We investigate the divergent perturbative series of  correlation functions for a massless, self-interacting scalar field in de Sitter space.  First, we use our previously proposed method of autonomous equations to obtain finite time-dependent functions, and show that these functions approximate the time evolution of the correlation functions of the stochastic approach reasonably well. Second, we apply the technique of autonomous equations to the Borel--Le Roy transforms of correlation functions, and use solutions of these equations to perform Borel resummation. The results match the time evolution obtained in the stochastic picture substantially better. In addition, we propose an alternative method for extracting perturbative coefficients and provide a new derivation of our autonomous equation by truncating a system of Schwinger--Dyson-type differential equations.}

\begin{document}

%\preprint{}


\maketitle

\section{Introduction}

In general, all calculations in quantum field theory can be divided into two classes. The~first includes perturbative methods, of which the most popular is the computation of Feynman diagrams. Attempts to extract additional information from perturbative results by performing some kind of resummation belong to the second class. The renormalization group formalism and various techniques for resummation of asymptotic series are very important examples of these investigations. 

The quantum field theoretical calculations in curved spacetime are more complicated and cumbersome than their counterparts in Minkowski spacetime. Quantum field theory on de Sitter background became very popular because this manifold possesses maximal symmetry. Thus, it represents a convenient arena in which to develop methods that in the future could be used in a wider context~\cite{dS,Tagirov,dS1,dS2,Allen,All-Fol,dS3,dS4,dS5,dS6,dS7}. On the other hand, de Sitter space represents a good approximation for the description of both the very early inflationary stage and the present late-time cosmic acceleration stage of our Universe. 

Remarkably, the development of quantum field theory in an expanding universe not only introduces more difficulties, but also offers some new opportunities that are absent in flat spacetime. The non-trivial causal structure of curved expanding spacetimes (for example, the Poincar\'e patch of de Sitter space) permits us to introduce natural frontiers between the long-wavelength and short-wavelength modes, and to focus our attention on the study of the dynamics of the former. The behavior of the long-wavelength modes is very interesting from a physical point of view because they represent the seeds from which the large-scale structure of the Universe forms. 

A non-perturbative approach for treating the long-wavelength (infrared) and short-wavelength (ultraviolet) modes differently was proposed by Starobinsky~\cite{Star} and further developed in many papers~\cite{Star-Yok,Finelli,Finelli1,Vennin}. Particularly important is the paper by Starobinsky and Yokoyama~\cite{Star-Yok}, where it was suggested that the dynamics of the long-wavelength part of a quantum scalar field in de Sitter space can be described by a classical stochastic variable whose probability satisfies a Fokker--Planck equation. At late times, any solution of this equation approaches a static solution, which can be used to calculate the late-time expectation values. In essence, Starobinsky's Fokker--Planck equation manages to resum the leading infrared logarithms of the perturbative expansion~\cite{Tsam-Wood}. The emergence of the stochastic picture from the full quantum theory was presented in later works \cite{Q-S,Q-S1,Q-S2,Tokuda,Pinol,Baumgart,Gorbenko}. A very clear explanation of the reasons why the stochastic formalism gives results coinciding with those derived from quantum field theory in the long-wavelength approximation is presented in~\cite{Woodard}. The Fokker--Planck equation is obtained as usual from the Langevin equation; see e.g.~\cite{Risken}. The Langevin equation for the long-wavelength modes was obtained in~\cite{Star} as a result of treating the short-wavelength modes as a source of the noise term. It is explained in \cite{Woodard} that the structure of the Langevin equation coincides with that of the truncated Yang--Feldman equation for the Heisenberg operators in quantum field theory~\cite{Yang-Feldman}, and the former can be obtained from the latter; see also \cite{Woodard0}. 

For calculations of cosmological correlators the Schwinger--Keldysh or ``in-in'' formalism~\cite{Schwinger, Keldysh, Bakshi, Bakshi1, Jordan, Collins} is used. The diagrams of the Schwinger--Keldysh technique have the same topological structure as their counterparts -- Feynman diagrams in the ``in-out'' formalism. However, in the Schwinger--Keldysh formalism, instead of a single vertex arising in a Feynman diagram, one should consider two vertices, (the so-called ``$+$'' and ``$-$'' vertices), and instead of one Feynman propagator one has to take into account four types of Green's functions. Obviously, all of this turns the calculation of the correlation functions by using the Schwinger--Keldysh formalism into a rather hard task; see, e.g.,~\cite{we,we1}. Remarkably, the results obtained within this formalism can also be obtained with the help of the iterative solution of the Yang--Feldman equation. Calculations of this type were performed in~\cite{Onemli, Karakaya-Onemli,Karakaya-Onemli1,we,we2,KP2026,Karakaya}. When considering the long-wavelength modes, the calculations in the Yang--Feldman formalism look much simpler, and instead of evaluating diagrams, one simply calculates a series of nested integrals. The correspondence between these two formalisms was described in some detail in~\cite{we2,Polina_thesis}.  

Once the equivalence between the quantum field theory formalism and the stochastic approach is well established, one can use whichever looks more suitable and less cumbersome. For example, in paper~\cite{we3}, the Fokker--Planck equation was used to construct the first-order differential equation establishing relations between different correlation functions. This permitted us to calculate, in a relatively simple way, various equal- and multi-time correlation functions. Our results coincided with those obtained earlier in quantum field theory~\cite{we2}. Anticipating the presentation of the content of the present paper, we can say that there is another simple method for computing the perturbative coefficients of the equal-time correlation functions. 

\newpage
As was mentioned above, the stochastic approach performs a resummation of the leading infrared logarithms of a quantum scalar field in de Sitter space. One of the most efficient methods of resummation of perturbative contributions in quantum field theory is the renormalization group approach; see, e.g.,~\cite{Bog-Shir, Shir}. While it was born in the context of the renormalization of the ultraviolet divergences in quantum field theory, later it became clear that its applications reside in a much broader area~\cite{Wil-Kog,Shir-Kov}. For example, in paper~\cite{Dyn-RG}, it was shown how one can improve the naive perturbative solutions of some rather complicated differential equations. This approach was called ``dynamical renormalization group method''. An interesting attempt to use this method for resummation of secular effects in de Sitter space was undertaken in paper~\cite{Holman-RG}. However, the results obtained in~\cite{Holman-RG} did not reproduce those known from the stochastic approach.  

In paper~\cite{we}, we developed a semi-heuristic method inspired by the renormalization group. The first terms of the perturbative expansion of the two-point and four-point functions for a massless scalar field with a quartic self-interaction in de Sitter space were used to construct autonomous differential equations. These equations were constructed in such a way that the perturbative results (up to a given order) can be obtained by simple iterations. If we take our correlation functions up to first order in the coupling constant, the exact solutions of the corresponding autonomous equations are simply those calculated in the Hartree--Fock approximation~\cite{Star-Yok}. If we take into account the second order perturbative term, we obtain a more complicated autonomous equation. One can look for its solution as a sum of the Hartree--Fock solution plus an additional smaller term. Functions obtained in this way are free from secular divergences and are well defined for all values of the cosmic time parameter. Moreover, their asymptotic values in the distant future are astonishingly close to those arising in the stochastic picture. In the subsequent paper~\cite{we2,KP2026}, the autonomous equation was constructed for the model of a massive scalar field with a quartic self-interaction. This equation looks much more cumbersome, but it can still be solved in the same approximation that was used in~\cite{we}. The asymptotic expressions for the correlation functions once again are close to those obtained by using the stochastic approach. The autonomous equation approach was also applied to the study of the behavior of different spectator fields on de Sitter background in~\cite{Bhat0, Bhat, Bhat1}. 

An attractive feature of the autonomous equation approach is that it gives explicit time-dependent functions for the correlation functions, and not only their asymptotic values in the distant future. In the stochastic approach, one does not have such functions. Indeed, we must numerically solve the Fokker-Planck equation for the probability density, and then we can calculate, also numerically, the time-dependent correlation functions. One of the goals of the present paper is to compare the analytic solutions of the autonomous equations with the results coming from the numerical integration of the Fokker--Planck equation. Remarkably, the corresponding evolutions indeed appear to be fairly close. This part of our present work was in some way motivated by paper~\cite{Borel-sec}. The authors of~\cite{Borel-sec} numerically calculate the time evolution of the correlation functions of the stochastic theory and compare them with those obtained by combining the use of a large number of coefficients of the perturbation theory with Pad\'e approximation and Borel resummation. Thus, the second goal of our paper is also partially inspired by~\cite{Borel-sec}. Namely, we combine the autonomous equations with the generalized Borel (Borel--Le Roy) resummation and compare the results with the correlation functions of the stochastic approach. Again, the agreement looks very good.

The third goal of this work is to give some additional arguments in favor of the autonomous equations, or, in other words, to give a new derivation of these equations, different from that presented in our paper~\cite{we}. Our approach to this task was influenced by~\cite{Schwinger-D}, where the authors studied the Schwinger--Dyson equations for the model with a massive scalar field and quartic self-interaction in a zero-dimensional field theory; see also works~\cite{Bender:1988bp,Okopinska:1990pt,Bender:2022eze,Bender:2023ttu} in the Euclidean zero-dimensional field theory~\cite{zero}. In this case, the general formula for all correlation functions of the theory is known and is expressed in terms of the modified Bessel (Macdonald) functions. These formulas have the same functional form as the asymptotic formulas arising in the stochastic approach (see~\cite{we1, we2, Beneke}). At the same time, there is an infinite system of algebraic Schwinger--Dyson equations connecting different $n$-point functions. By truncating this system, the authors of~\cite{Schwinger-D} found its solution in terms of rational approximants. Here, we construct an analogous system of Schwinger--Dyson-type equations for a one-dimensional field theory (we add one temporal dimension). In this case, the corresponding equations are not algebraic, but rather first-order differential equations that coincide with the recursion relations arising in the stochastic approach~\cite{Tsam-Wood, Q-S, we3}. Then we consider different types of truncations of the obtained system. If we set to zero all $n$-point functions with $n$ bigger than some number, we obtain a closed system of equations, which can be solved explicitly in terms of elementary functions. Expanding these functions in power series with respect to the time parameter gives a certain number of coefficients that coincide with those found by other field-theoretical methods. However, we can try to implement another type of truncation by using the Gaussian approximation for the six-point correlation function. In this case, we obtain a closed system of equations for the two-point and four-point correlation functions. Studying this system leads to our autonomous equations for the two-point and four-point functions. 

The paper is organized as follows. In section~\ref{autonom_eqs}, we briefly review the formalism of autonomous equations. In section~\ref{numerical}, we apply this formalism to the theory of a massless scalar field with a quartic self-interaction in de Sitter space, and compare the solutions of our autonomous equations with the equal-time two-point and four-point functions obtained from the numerical solution of the Fokker--Planck equation. In sections~\ref{Borel-LeRoy} and~\ref{singularities}, we combine the technique of Borel--Le Roy resummation with the implementation of autonomous equations, and compare the obtained results with the time evolution of correlation functions in the stochastic picture. In section~\ref{singularities}, we also briefly discuss the singularities of Borel transforms. The last section~\ref{S_D_eqs} is devoted to the Schwinger--Dyson-type equations and different schemes of their truncation. Section~\ref{conclusion} contains concluding remarks.

\section{Autonomous equations}\label{autonom_eqs}

Let us suppose that we would like to study a certain function $f(t)$, and that we only know the first terms of its expansion with respect to a small parameter $\lambda$, this expansion being simultaneously an expansion in the time parameter $t$:
\begin{equation}
f(t) = At - \lambda B t^3 + \lambda^2C t^5 + {\cal O}(\lambda^3). 
\label{auton}\end{equation}  
Our suggestion in the paper~\cite{we} was the following: we would like to find an autonomous first-order differential equation that produces the terms of the expansion in~\eqref{auton} by an iterative procedure. Let us start with the expression~\eqref{auton}, where we only keep the term that is zeroth order in~$\lambda$: 
%\begin{equation}
$f(t) = At$. 
%\label{auton1} \end{equation}
This function can be obtained as a solution of a simple first-order differential equation, 
%\begin{equation}
%\frac{df}{dt} 
$\dot{f}= A$. Here and in the following, the dot denotes the derivative with respect to time.
%\label{auton2} \end{equation}
If we take into account the first two terms in the expansion~\eqref{auton}, then we have the following equation:
\begin{equation}
%\frac{df}{dt} 
\dot{f} = A- \frac{3\lambda B}{A^2} \, f^2.
\label{auton3}
\end{equation}
Equation \eqref{auton3} has a simple, exact solution:
\begin{equation}
f(t) = \sqrt{\frac{A^3}{3\lambda B}}\tanh\left(\sqrt{\frac{3\lambda B}{A}}t\right),
\label{auton4}
\end{equation}
where the initial condition is chosen as 
%\begin{equation}
$f(0) = 0$. 
%\label{initial} \end{equation}
Expanding function~\eqref{auton4} in the small $\lambda$ regime reproduces the first two terms of~\eqref{auton}.
At the same time, this function, in contrast to expansion~\eqref{auton}, is free of secular divergences and is well-defined for all values of $t$. %Moreover, it has a non-analytic dependence on the small parameter $\lambda$. 
Analogously, to reproduce all three terms in expansion~\eqref{auton}, we write down the following autonomous equation:
\begin{equation}
 %\frac{df}{dt} 
 \dot{f} = A- \frac{3\lambda B}{A^2}f^2 +\lambda^2\left(\frac{5C}{A^4}-\frac{6B^2}{A^5}\right)f^4.
\label{auton5}
\end{equation}
In general, it is not possible to find the explicit form of $f(t)$ that satisfies this equation. However, sometimes one can obtain a solution in the form of a perturbative expansion in a small parameter, describing the deviation of the solution of equation~\eqref{auton5} from that of equation~\eqref{auton3}. Introducing the parameter 
%\begin{equation}
$\epsilon\equiv\dfrac{5AC}{6B^2}-1$, 
%\label{epsilon}\end{equation}
we can rewrite our equation~\eqref{auton5} as follows:
\begin{equation}
%\frac{df}{dt} 
\dot{f}= A- \frac{3\lambda B}{A^2}\, f^2 + \frac{6 \lambda^2 B^2 \epsilon}{A^5} \, f^4.
\label{auton6}
\end{equation}
We can look for its solution in the form 
%\begin{equation}
$f(t) = f_0(t)+\epsilon f_1(t)$, 
%\label{auton7}\end{equation}
where $f_0(t)$ is the solution of equation~\eqref{auton3}, given by formula \eqref{auton4}, while $f_1(t)$ satisfies the equation
\begin{equation}
%\frac{df_1}{dt} 
\dot{f}_1= - \frac{6\lambda B}{A^2} \, f_0f_1+\frac{6\lambda^2B^2}{A^5}\,f_0^4.
\label{auton8}
\end{equation}
After solving this equation, we find~\cite{we}
%\begin{widetext}
\begin{equation}
f(t) = \biggl(1+\frac{\epsilon}{3}\biggr)\sqrt{\frac{A^3}{3\lambda B}}\tanh\left(\sqrt{\frac{3\lambda B}{A}} \, t\right)+
\frac{\epsilon\left(\dfrac{2}{3}\sqrt{\dfrac{A^3}{3\lambda B}}\tanh\left(\sqrt{\dfrac{3\lambda B}{A}} \,t\right)-At\right)}{\cosh^2\left(\sqrt{\dfrac{3\lambda B}{A}} \, t\right)}.
\label{auton9}
\end{equation}
%\end{widetext}
At late times, $t \rightarrow \infty$, this expression tends to 
\begin{equation}
f(t) \rightarrow \sqrt{\frac{A^3}{3\lambda B}}\left(\frac23+\frac{5}{18}\frac{AC}{B^2}\right).
\label{auton10}
\end{equation}

One can also construct autonomous equations and find their solutions for an expansion different from~\eqref{auton}. We shall consider the following expansion:
\begin{equation}
g(t) = Jt^2-\lambda Kt^4 +\lambda^2 L\,t^6+ {\cal O}(\lambda^3).
\label{auton11}
\end{equation}
If we only keep the first two terms on the r.h.s. of~\eqref{auton11}, the autonomous equation~is 
\begin{equation}
%\frac{dg}{dt} 
\dot{g} = 2\sqrt{Jg} - \frac{3\lambda K}{J^{3/2}} \, g^{3/2}, 
\label{auton12}
\end{equation}
and its solution is 
\begin{equation}
g(t) = \frac{2J^2}{3\lambda K}\tanh^2\left(\sqrt{\frac{3\lambda K}{2J}}t\right).
\label{auton13}
\end{equation}
The autonomous equation corresponding to all three terms in expansion~\eqref{auton11} is
\begin{equation}
%\frac{dg}{dt} 
\dot{g} = 2\sqrt{Jg}- \frac{3\lambda K}{J^{3/2}} \, g^{3/2} +\lambda^2\left(\frac{5L}{J^{5/2}}-\frac{17}{4}\frac{K^2}{J^{7/2}}\right)g^{5/2}.
\label{auton14}
\end{equation}
By defining the parameter
%\begin{equation}
$\tilde{\epsilon} \equiv\dfrac{20JL}{17K^2}-1$
%\label{epsilon1}\end{equation}
and proceeding similarly to the previous example, one finds the solution, with $g(0)=0$,  in the following form:
%\begin{widetext}
%\begin{equation}
\begin{align}\label{auton15}
g(t) & ={2J^2\over3\lambda K}\biggl(1+\frac{17}{18}\tilde{\epsilon}\biggr)\tanh^2\left(\sqrt{\frac{3\lambda K}{2J}} \, t\right) \\ \nonumber 
& \quad +\frac{34}{27}\frac{J^2\,\tilde{\epsilon}}{\lambda K} \frac{\tanh\left(\sqrt{\dfrac{3\lambda K}{2J}}t\right)}{\cosh^2\left(\sqrt{\dfrac{3\lambda K}{2J}}t\right)}\left(\tanh\left(\sqrt{\frac{3\lambda K}{2J}}t\right)-\frac{3}{2}\sqrt{\frac{3\lambda K}{2J}}t\right).
\end{align} %\end{equation}
%\end{widetext}
When $t \rightarrow \infty$, one has
\begin{equation}
g(t) \rightarrow \frac{2J^2}{3\lambda K}\left(\frac{1}{18}+\frac{20}{18}\frac{JL}{K^2}\right).
\label{auton16} \end{equation}

\vspace{0.1cm}
\section{Autonomous equations for correlation functions in de Sitter space vs. the numerical solution of the Fokker{--}Planck equation}\label{numerical}

In this section, we are going to apply the technique of autonomous equations for resummation of secular terms in de Sitter space and compare the solutions to these equations 
with the correlation functions obtained by numerically solving the Fokker--Planck equation. 

Let us consider de Sitter space in the expanding flat coordinates with the following metric:
\begin{equation}
ds^2 = dt^2 -e^{2Ht}\delta_{ij}dx^idx^j,
\label{dS} \end{equation}
where $H$ is the Hubble constant or the inverse of the de Sitter radius. We have a massless spectator scalar field  with a quartic self-interaction, living in this space:
\begin{equation}
S = \int d^4x\sqrt{-g}\left(\frac12 \, g^{\mu\nu}\partial_{\mu}\phi \partial_{\nu}\phi-\frac{\lambda}{4} \,\phi^4\right). 
\label{action} \end{equation}

At the level of the free theory, the long-wavelength part of the two-point function at coinciding space-time points has the form~\cite{Star-s,Linde,Vil-Ford,Ford-Vil,All-Fol}:
%\begin{equation}
$\bigl\langle \phi^2 (t,\x\,) \bigr\rangle= \dfrac{H^3 t}{4\pi^2}$, 
%\label{sec} \end{equation} 
and shows linear secular growth. 
Taking into account the self-interaction of the scalar field yields 
\begin{equation}
\bigl\langle \phi^2 (t,\x\,) \bigr\rangle = \frac{H^3 t}{4\pi^2} - \frac{\lambda H^5 t^3}{24\pi^4} + \frac{\lambda^2 H^7 t^5}{80\pi^6} + {\cal O} \bigl(\lambda^3\bigr).
\label{sec1}
\end{equation}
The coefficients in~\eqref{sec1} were obtained by various methods~\cite{Tsam-Wood,we,Onemli,Onem-Wood,Karakaya-Onemli,Brunier,Kahya,we3}.
Apparently, the structure of expression~\eqref{sec1} coincides with that of~\eqref{auton}, and the coefficients are
\begin{equation}
A = \frac{H^3}{4\pi^2},\qquad B = \frac{H^5}{24\pi^4},\qquad \text{and} \qquad C = \frac{H^7}{80\pi^6}.
\label{coeff}
\end{equation}
Hence, we have solution \eqref{auton9} of equation~\eqref{auton6} with the above coefficients~\cite{we}: 
%\begin{widetext}
\begin{equation}
\bigl\langle \phi^2 (t,\x\,) \bigr\rangle = \frac{7H^2}{12\sqrt{2\lambda}\, \pi}\tanh\left(\sqrt{\frac{H^2\lambda}{2\pi^2}}t\right)+
\frac{\dfrac{H^2}{6\sqrt{2\lambda}\pi}\tanh\left(\sqrt{\dfrac{H^2\lambda}{2\pi^2}}t\right)-\dfrac{H^3t}{8\pi^2}}{\cosh^2\left(\sqrt{\dfrac{H^2\lambda}{2\pi^2}}t\right)}.
\label{sec2}
\end{equation}
%\end{widetext}
Note that solution~\eqref{auton4} to equation~\eqref{auton3} with coefficients~\eqref{coeff} coincides with the long-wavelength part of $\bigl\langle \phi^2 (t,\x\,) \bigr\rangle$ in the Hartree--Fock approximation~\cite{Star,Star-Yok,we}. In the late-time limit, $t \rightarrow \infty$, function~\eqref{sec2} approaches
\begin{equation}
\bigl\langle \phi^2 (t,\x\,) \bigr\rangle \rightarrow \frac{7H^2}{12\sqrt{2\lambda}\pi}\approx \, 0.13130 \, \frac{H^2}{\sqrt{\lambda}},
\label{sec3} \end{equation}
which is close to the asymptotic value arising in the Starobinsky--Yokoyama stochastic approach~\cite{Star,Star-Yok}. 

The stochastic approach~\cite{Star,Star-Yok} matches the long-wavelength part of the quantum scalar field $\phi(t,\vec{x})$ to the classical stochastic field $\varphi$ with the probability distribution function ${\cal P}(\varphi,t)$ that satisfies the Fokker--Planck equation,
\begin{equation}
\frac{\partial {\cal P}}{\partial t} = \frac{1}{3H}\frac{\partial}{\partial \varphi}\left(\frac{\partial V}{\partial \varphi}{\cal P}\right)+\frac{H^3}{8\pi^2}\frac{\partial^2 {\cal P}}{\partial \varphi^2}, 
\label{F-P}
\end{equation} 
and the time-dependent expectation values given by
\begin{equation}
\langle \varphi^{n}\rangle(t)\equiv\int_{-\infty}^{\infty} d\varphi \, \varphi^{n} \, {\cal P}(\varphi,t).
\label{S-D3} \end{equation}
Here $V(\varphi)$ is the same potential that one investigates in the quantum field theory model. At late times, any solution of equation~\eqref{F-P}, with the quartic potential $V=\dfrac{\lambda}{4}\varphi^4$, approaches the static solution~\cite{Star-Yok}:
\begin{equation}\label{static_prob}
{\cal P}_{\infty}(\varphi) = \left(\frac{32\pi^2\lambda}{3}\right)^{\frac14}\frac{1}{\Gamma\left(\frac14\right)H}\exp\left(-\frac{2\pi^2\lambda\varphi^4}{3H^4}\right)\;.
\end{equation}
Using this probability distribution function, one can calculate~\cite{Star-Yok} the asymptotic value of $\langle \varphi^{2}\rangle(t)$ via~\eqref{S-D3}:
 \begin{equation}
\langle \varphi^2\rangle(t)\rightarrow\sqrt{\frac{3}{2\pi^2}}\frac{\Gamma\left(\frac34\right)H^2}{\Gamma\left(\frac14\right)\sqrt{\lambda}}\, \approx \, 0.13176 \, \frac{H^2}{\sqrt{\lambda}}.
\label{sec4}
\end{equation}
As we see, the relative difference between values \eqref{sec3} and \eqref{sec4} is only $\approx 0.36 \%$~\cite{we}.
  
Let us also consider the perturbative structure of the long-wavelength part of $\bigl\langle \phi^4 (t,\x\,) \bigr\rangle$, which behaves as follows~\cite{Tsam-Wood,we,we2,we3,Bhat1}:
\begin{equation}
\bigl\langle \phi^4 (t,\x\,) \bigr\rangle = \frac{3H^6 t^2}{16\pi^4}- \frac{3\lambda H^8t^4}{32\pi^6} + \frac{53\lambda^2 H^{10} t^6 }{960\pi^8} + {\cal O} \bigl(\lambda^3\bigr).
\label{sec5} \end{equation}
It has the form of our expression~\eqref{auton11} with the coefficients:
\begin{equation}
J = \frac{3H^6}{16\pi^4},\qquad  K = \frac{3H^8}{32\pi^6},\qquad \text{and} \qquad L = \frac{53H^{10}}{960\pi^8}.
\label{coeff1} \end{equation}
We already know the solution to the corresponding autonomous equation, see~\eqref{auton15},
%\begin{widetext}
%\begin{equation}
\begin{align} \label{sec6}
\bigl\langle \phi^4 (t,\x\,) \bigr\rangle & = \frac{221}{648\pi^2 }\frac{H^4}{\lambda}\tanh^2\left(\sqrt{\frac{3\lambda}{4\pi^2}}Ht\right) \\ \nonumber 
& \quad 
+\frac{59}{324\pi^2}\frac{H^4}{\lambda} \frac{\tanh\left(\sqrt{\dfrac{3\lambda}{4\pi^2}}Ht\right)}{\cosh^2\left(\sqrt{\dfrac{3\lambda}{4\pi^2}}Ht\right)}\left(\tanh\left(\sqrt{\frac{3\lambda}{4\pi^2}}Ht\right)-\frac{3}{2}\sqrt{\frac{3\lambda}{4\pi^2}}Ht\right),
\end{align} %\end{equation}
%\end{widetext}
and its asymptotic value for $t \rightarrow \infty$ is 
\begin{equation}
\bigl\langle \phi^4 (t,\x\,) \bigr\rangle \rightarrow \, \frac{221 H^4}{648\pi^2 \lambda} \approx \, 0.03456 \, \frac{H^4}{\lambda}.
\label{sec7}
\end{equation}
The asymptotic value for the corresponding quantity in the stochastic approach is 
\begin{equation}
\langle \varphi^4\rangle(t)\rightarrow\frac{3H^4}{8\pi^2\lambda}\approx0.03800 \, \frac{H^4}{\lambda}.
\label{sec8} \end{equation}
The agreement between values~\eqref{sec7} and~\eqref{sec8}  is worse than between~\eqref{sec3} and~\eqref{sec4}, but it is still fairly good: their relative difference is $\approx 9 \%$.

In order to compare functions~\eqref{sec2} and~\eqref{sec6} with the time evolution of $\langle \varphi^2\rangle(t)$ and $\langle \varphi^4\rangle(t)$ in the stochastic picture, it is convenient to rescale our variables: 
\begin{equation}
\bar\phi\equiv{\lambda^{1/4}\over H}\phi, \qquad 
\label{resphi}
%\end{equation} \begin{equation}
\bar\varphi\equiv{\lambda^{1/4}\over H}\varphi, \qquad \text{and} \qquad 
%\label{resvarphi}\end{equation}\begin{equation}
\bar N\equiv\lambda^{1/2}N,
%\label{resN} 
\end{equation}
where $N=Ht$ is the number of $e$-folds, defined as the logarithm of the scale factor. With these new variables, our solutions~\eqref{sec2} and~\eqref{sec6} can be rewritten as
%\begin{eqnarray}
\begin{align}
\left\langle\bar \phi^2\right\rangle(\bar N) &= \frac{7\sqrt2}{24\pi}\tanh\left(\frac{\bar N}{\sqrt2\pi}\right)+\frac{\tanh\left(\dfrac{\bar N}{\sqrt2\pi}\right)-\dfrac{3\bar N}{2\sqrt2\pi}}{6\sqrt2\pi\cosh^2\left(\dfrac{\bar N}{\sqrt2\pi}\right)\, }\, , %\hspace{-0.5cm} \nonumber \\
\label{restwop} \\
%\end{eqnarray} and \begin{widetext} \begin{equation}
\label{resfourp}
\left\langle\bar \phi^4\right\rangle(\bar N) &= \frac{221}{648\pi^2}\tanh^2\left(\frac{\sqrt3\bar N}{2\pi}\right) %\\ \nonumber & \qquad 
+\frac{59}{216\pi^2}\frac{\tanh\left(\dfrac{\sqrt3\bar N}{2\pi}\right)}{\cosh^2\left(\dfrac{\sqrt3\bar N}{2\pi}\right)}\left(\dfrac23\tanh\left(\frac{\sqrt3\bar N}{2\pi}\right)-\frac{\sqrt3\bar N}{2\pi}\right)\,.
\end{align} %\end{equation}
%\end{widetext}
The Fokker-Planck equation \eqref{F-P}, with the quartic potential, can also be rewritten in terms of rescaled variables \eqref{resphi}:
\begin{equation}
\frac{\partial {\cal P}}{\partial\bar N} = \frac{1}{3}\frac{\partial}{\partial \bar\varphi}\left({\bar\varphi^3\cal P}\right)+\frac{1}{8\pi^2}\frac{\partial^2 {\cal P}}{\partial\bar\varphi^2}. 
\label{FPres}
\end{equation} 
Our new functions~\eqref{restwop} and \eqref{resfourp}, as well as the Fokker--Planck equation~\eqref{FPres}, have no explicit dependence on $H$ and $\lambda$. Hereby, we do not have to worry about specific values of these parameters. 

Since solutions~\eqref{restwop} and~\eqref{resfourp}, as well as the original perturbative series~\eqref{sec1} and~\eqref{sec5}, are equal to zero at $\bar N=0$ (or $t=0$), we assume the initial condition for the Fokker--Planck equation~\eqref{FPres} to be ${\cal P}(\bar\varphi,\bar N=0)=\delta(\bar\varphi)$. To solve~\eqref{FPres} numerically, we follow~\cite{Borel-sec} and use the narrow Gaussian function, \mbox{${\cal P}(\bar\varphi,\bar N=0)=\dfrac{e^{-\tfrac{\bar\varphi^2}{2(0.02)^2}}}{(0.02\sqrt{2\pi})}$}, to mimic the $\delta$-function.  

\begin{figure}[htbp]
    \centering
    \includegraphics[scale=0.52]{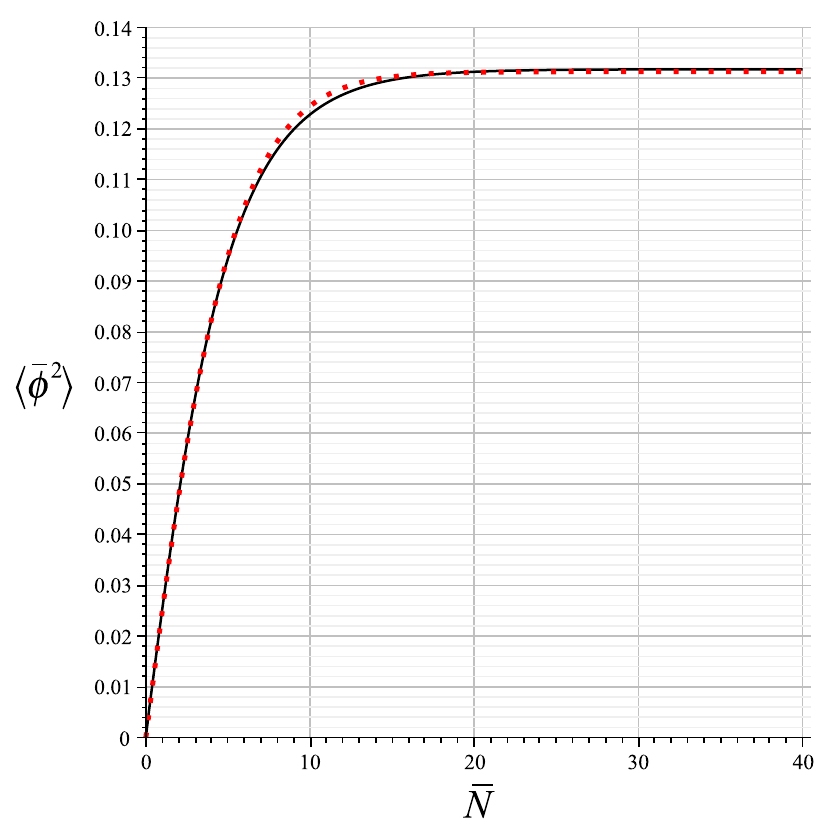}\hfill
    \includegraphics[scale=0.52]{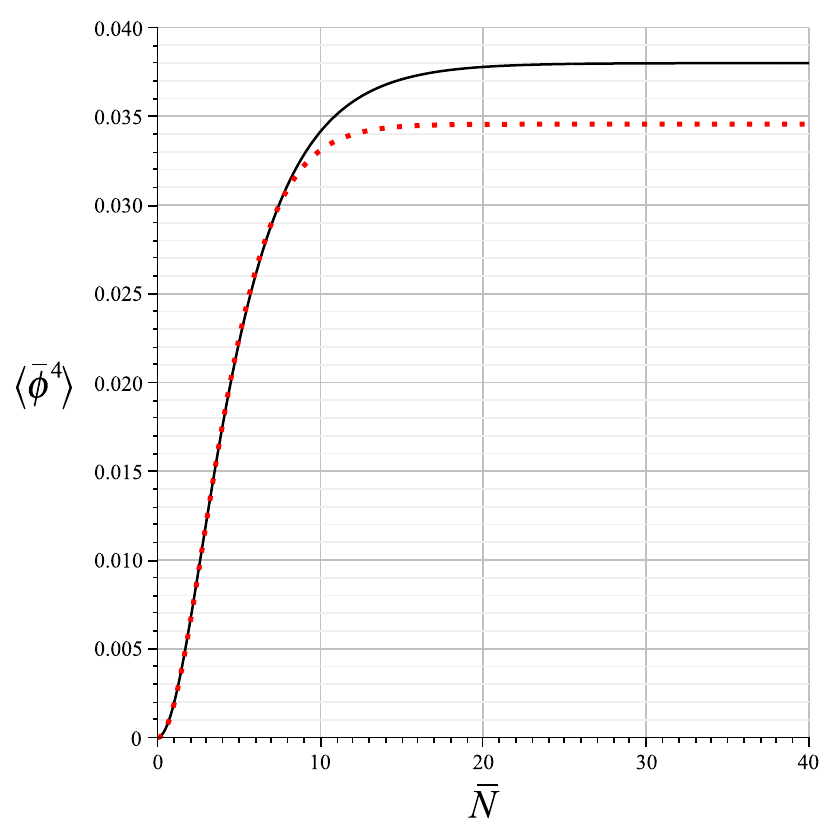}
    \caption{Left: Comparison of $\left\langle\bar \phi^2\right\rangle(\bar N)$ and $\left\langle\bar \varphi^2\right\rangle(\bar N)$; the red dotted line is the plot of function \eqref{restwop} obtained from the autonomous equation, while the black solid line corresponds to $\langle\bar\varphi^2\rangle(\bar N)$, which is calculated from the numerical solution ${\cal P}(\bar\varphi,\bar N)$ of the Fokker{--}Planck  equation \eqref{FPres}. Right: Comparison of $\left\langle\bar \phi^4\right\rangle(\bar N)$ and $\left\langle\bar \varphi^4\right\rangle(\bar N)$; the red dotted line is the plot of function \eqref{resfourp}, and the black solid line corresponds to $\langle\bar\varphi^4\rangle(\bar N)$, which is calculated from the numerical solution ${\cal P}(\bar\varphi,\bar N)$ of the Fokker{--}Planck  equation \eqref{FPres}.}
    \label{fig:both}
\end{figure}

By looking at %Figure~\ref{fig:one} and Figure~\ref{fig:two}
Figure~\ref{fig:both}, we see that, qualitatively, the behavior of both correlation functions is reproduced quite well by the solutions to the autonomous equations. The quantitative agreement between $\left\langle\bar \phi^2\right\rangle(\bar N)$ and $\left\langle\bar \varphi^2\right\rangle(\bar N)$ is also good: the biggest relative difference between them occurs around $\bar N\approx9$ and equals $\approx1.4\%$. In the case of $\left\langle\bar \phi^4\right\rangle(\bar N)$ and $\left\langle\bar \varphi^4\right\rangle(\bar N)$, the quantitative accuracy is worse: they start to diverge from each other at around $\bar N\approx8$, and evolve towards the respective asymptotic values, which, as mentioned earlier, have a relative difference $\approx9\%$.  
%\begin{figure}
%    \begin{center}
%        \includegraphics[scale=0.6]{2point_fine.pdf}
%        \caption{Comparison of $\left\langle\bar \phi^2\right\rangle(\bar N)$ and $\left\langle\bar \varphi^2\right\rangle(\bar N)$: the red dotted line is the plot of function \eqref{restwop} obtained from the autonomous equation, and the black solid line corresponds to $\langle\bar\varphi^2\rangle(\bar N)$, which is calculated from the numerical solution ${\cal P}(\bar\varphi,\bar N)$ of the Fokker-Planck  equation \eqref{FPres}.}\label{fig:one}
%\end{center} \end{figure}
%\begin{figure}\begin{center}
%        \includegraphics[scale=0.6]{4point.pdf}
%        \caption{Comparison of $\left\langle\bar \phi^4\right\rangle(\bar N)$ and $\left\langle\bar \varphi^4\right\rangle(\bar N)$: the red dotted line is the plot of function \eqref{resfourp} obtained from the autonomous equation, and the black solid line corresponds to $\langle\bar\varphi^4\rangle(\bar N)$, which is calculated from the numerical solution ${\cal P}(\bar\varphi,\bar N)$ of the Fokker-Planck  equation \eqref{FPres}.}\label{fig:two}
%     \end{center}
%\end{figure}

\section{Autonomous equations and Borel{--}Le Roy resummation}\label{Borel-LeRoy}

The authors of~\cite{Borel-sec} consider the perturbative series for correlation functions in the model of a massless scalar field with a quartic interaction in de Sitter space. These are the same as series~\eqref{sec1} and \eqref{sec5}, but with a huge number of coefficients. Indeed, the general formula for the coefficients of the expansion of the two-point function is presented in~\cite{Borel-sec}, but, unfortunately, its derivation is not explained. By taking the truncated series, the authors of~\cite{Borel-sec} calculate its Borel transform, then use the transform's Pad\'e approximations to perform Borel resummation by the Laplace transformation. The resulting functions are compared with the correlation functions obtained from the numerical solution of the Fokker{--}Planck equation~\eqref{F-P}. For large enough orders of the Pad\'e approximants, and a certain hierarchy between the orders of their numerators and denominators, the Borel resummation correctly reproduces the time evolution of the correlation functions.    

In this section, we intend to combine our autonomous equations with the technique of Borel resummation in an unusual way. Let us suppose that we have a divergent series with the following structure: 
\begin{equation}
f(t) = \sum_{n=0}^{\infty}a_nt^n.
\label{series}
\end{equation}
Note that our series for $\bigl\langle \phi^2 (t,\x\,) \bigr\rangle$ and $\bigl\langle \phi^4 (t,\x\,) \bigr\rangle$ have this structure. The Borel transform of this series~\eqref{series} is defined as~\cite{Borel,Hardy,Bender,Suslov,Fantini:2024ctk}:
\begin{equation}
f_{\rm B}(z) = \sum_{n=0}^{\infty}\frac{a_n}{\Gamma(n+1)} \, z^n.
\label{series1}
\end{equation}
This series is more convergent, and in some cases it is possible to find a good analytical continuation 
%\begin{equation}
$f_{\rm B}(z) \rightarrow \tilde{f}_{\rm B}(z)$. 
%\label{series2} \end{equation} 
Then the Borel resummation of $f(t)$ is defined by the Laplace transformation
\begin{equation}
f_{\rm S}(t) = \int_{0}^{\infty}dz \, e^{-z}\tilde{f}_{\rm B}(zt).
\label{series3}
\end{equation}
If we expand $\tilde{f}_{\rm B}(zt)$ around $t=0$ and perform integration~\eqref{series3} term by term, we arrive at the original series~\eqref{series}. Moreover, if there are no poles on the integration contour, then, unlike the divergent series~\eqref{series}, the Borel resummation $f_{\rm S}(t)$ has a finite value for finite non-zero values of $t$. 

In~\cite{Borel-sec}, the series expansions of the correlation functions are truncated at {\it fairly  high} finite orders, from which the Borel transformations are constructed, and Pad\'e approximants play the role of $\tilde{f}_{\rm B}(z)$. 

Here, we shall truncate the perturbative series at a {\it much lower} order, with only the first three terms being retained in the expansions of the corresponding correlation functions. Instead of the Borel transformations~\eqref{series1}, we shall use a more general Borel--Le Roy transformation~\cite{Hardy,Suslov,Peng-Shu},
\begin{equation} 
f_{\rm B}(z) = \sum_{n=0}^{\infty}\frac{a_n}{\Gamma(n+b+1)} \, z^n,
\label{series4}
\end{equation}
where $b$ is a free real parameter. We shall take the expansion~\eqref{series4} with three terms, construct the corresponding autonomous equation, as well as its linearized version, and solve it. Finally, to obtain the Borel resummation, we shall use the generalized version of transformation~\eqref{series3}, which has the form
\begin{equation}
f_{\rm S}(t) = \int_{0}^{\infty}dz z^be^{-z}\tilde{f}_{\rm B}(zt),
\label{series5}
\end{equation}
where the role of $\tilde{f}_{\rm B}(z)$ will be played by the solution of the autonomous equation. The obtained representations for the correlation functions will be compared with those coming from the numerical solution of the Fokker{--}Planck equation. 

The function $f_{\rm S}(t)$ in~\eqref{series5} depends on the free parameter $b$. This situation is rather typical for the problems where resummations of divergent series are considered; see, e.g.,~\cite{Suslov,Yukalov,Kaz-Shir-Tar}. In a way, it resembles the problem of the choice of some parameters in the process of renormalization of ultraviolet divergences. Here, we propose to fix the parameter $b$ as follows: we shall calculate the asymptotic value of function~\eqref{series5} at late times, $t \rightarrow \infty$, and set it equal to the known asymptotic value of the corresponding correlation function in the stochastic picture.

\vspace{0.1cm}
Before presentation of numerical results, let us write down all the necessary analytical expressions. The first three terms in the Borel{--}Le Roy transform of the correlation function~\eqref{sec1} are
%\begin{eqnarray}
\begin{equation}
\left\langle \phi^2\right\rangle_{\rm B}(z) = \frac{H^3z}{4\pi^2\, \Gamma(b+2)} - \frac{\lambda \, H^5z^3}{24\pi^4 \, \Gamma(b+4)} +\frac{\lambda^2H^7z^5}{80\pi^6 \, \Gamma(b+6)} + {\cal O} \bigl(\lambda^3\bigr).
\label{Boftwo}
\end{equation} %\end{eqnarray}
The solution to the corresponding autonomous equation has the same form as~\eqref{auton9}, with the appropriate coefficients,
%\begin{widetext} \begin{eqnarray}
\small{\begin{align} \label{autonB2}
&\left\langle \phi^2\right\rangle_{\rm B}(z) = \frac{H^2}{\sqrt{2\lambda}}\frac{\sqrt{(b+3)(b+2)}}{3\pi \, \Gamma(b+2)}\left(1+\frac{3(b+3)(b+2)}{4(b+5)(b+4)}\right)\tanh\left({\frac{\sqrt\lambda Hz}{\sqrt{2(b+3)(b+2)}\pi}}\right) \\ \nonumber 
& +\frac{H^2}{\sqrt{2\lambda}}\frac{\sqrt{(b+3)(b+2)}}{2\pi\Gamma(b+2)}\left(\frac{3(b+3)(b+2)}{2(b+5)(b+4)}-1\right) 
\frac{\left(\dfrac{2}{3}\tanh\left({\dfrac{\sqrt\lambda Hz}{\sqrt{2(b+3)(b+2)}\pi}}\right)-{\dfrac{\sqrt\lambda Hz}{\sqrt{2(b+3)(b+2)}\pi}}\right)}{\cosh^2\left({\dfrac{\sqrt\lambda Hz}{\sqrt{2(b+3)(b+2)}\pi}}\right)},
\end{align}} %\end{eqnarray}
%\end{widetext}
\normalsize 
and can be  used to obtain the Borel resummed correlation function, 
\begin{equation}
\left\langle \phi^2\right\rangle_{\rm S}(t) = \int_{0}^{\infty}dz z^be^{-z}\left\langle \phi^2\right\rangle_{\rm B}(zt).
\label{twopB}
\end{equation}
The integral in \eqref{twopB} cannot be expressed in terms of known functions, however, we can find its asymptotic value for $t \rightarrow \infty$. Indeed, as $t \rightarrow \infty$ we have
\begin{eqnarray} %\hspace{-4mm}
\left\langle \phi^2\right\rangle_{\rm B}(zt)\to \frac{H^2}{\sqrt{2\lambda}}\frac{\sqrt{(b+3)(b+2)}}{3\pi\Gamma(b+2)}\left(1+\frac{3(b+3)(b+2)}{4(b+5)(b+4)}\right).%\hspace{-1cm}\nonumber\\
\label{autonB3}
\end{eqnarray}
By substituting this expression into the integral in~\eqref{twopB}, we obtain
\begin{eqnarray}%\hspace{-4mm}
\left\langle \phi^2\right\rangle_{\rm S}(t)\to \frac{H^2}{\sqrt{2\lambda}}\frac{\sqrt{(b+3)(b+2)}}{3\pi(b+1)}\left(1+\frac{3(b+3)(b+2)}{4(b+5)(b+4)}\right). %\hspace{-1cm}\nonumber\\
\label{autonB4}
\end{eqnarray}
If we set expression~\eqref{autonB4} equal to the asymptotic value~\eqref{sec4} in the stochastic picture, we can fix the value of the parameter $b$. The obtained algebraic equation has three roots: two of them are complex numbers, and one is real and positive with the value $b \approx 13.55$. Only the latter one is physically relevant. 

We shall carry out the same procedure for the perturbative expansion~\eqref{sec5}. The first three terms in its Borel--Le Roy transform are 
%\begin{eqnarray}
\begin{equation}
\left\langle \phi^4\right\rangle_{\rm B}(z) = \frac{3H^6z^2}{16\pi^4\,\Gamma(b+3)}-\frac{3\lambda H^8z^4}{32\pi^6\,\Gamma(b+5)}+\frac{53\lambda^2H^{10}z^6}{960\pi^8\,\Gamma(b+7)} + {\cal O}\bigl(\lambda^3\bigr),
\label{Boffour} \end{equation} %\end{eqnarray}
and the solution to the corresponding autonomous equation reads, see~\eqref{auton15},
%\begin{widetext}
%\begin{eqnarray}
\begin{align}\label{autonB6}
\left\langle \phi^4\right\rangle_{\rm B}(z) & =\frac{H^4}{\lambda}\frac{{(b+3)(b+4)}}{72\pi^2\Gamma(b+3)}\left(1+\frac{212(b+3)(b+4)}{9(b+5)(b+6)}\right)\tanh^2\left({\frac{\sqrt{3\lambda} Hz}{2\pi\sqrt{(b+3)(b+4)}}}\right) \nonumber \\
&\quad +\frac{H^4}{\lambda}\frac{{17(b+3)(b+4)}}{24\pi^2\Gamma(b+3)}\left(\frac{212(b+3)(b+4)}{153(b+5)(b+6)}-1\right)\tanh\left({\frac{\sqrt{3\lambda} Hz}{2\pi\sqrt{(b+3)(b+4)}}}\right) \nonumber \\
&\qquad\times \frac{\dfrac23\tanh\left({\dfrac{\sqrt{3\lambda} Hz}{2\pi\sqrt{(b+3)(b+4)}}}\right)-{\dfrac{\sqrt{3\lambda} Hz}{2\pi\sqrt{(b+3)(b+4)}}}}{\cosh^2\left({\dfrac{\sqrt{3\lambda} Hz}{2\pi\sqrt{(b+3)(b+4)}}}\right)}.
\end{align} %\end{eqnarray}
%\end{widetext}
The Borel resummed correlation function $\left\langle \phi^4\right\rangle_{\rm S}(t)$ can be computed from~\eqref{autonB6} analogously to relation~\eqref{twopB}. As $t \rightarrow \infty$, its value approaches the following expression, 
%\begin{eqnarray} \hspace{-4mm}
\begin{equation}
\left\langle \phi^4\right\rangle_{\rm S}(t)\to \frac{\,\, H^4}{72\pi^2\lambda}\frac{{(b+3)(b+4)}}{(b+1)(b+2)}\left(1+\frac{212(b+3)(b+4)}{9(b+5)(b+6)}\right).
\end{equation} %\end{eqnarray}
By equating this expression to the asymptotic value~\eqref{sec8} in the stochastic picture, we obtain four values for the parameter $b$. Three of them are real numbers smaller than $-1$, therefore, for these values of $b$, the Laplace transformation \eqref{series5} does not converge around $z=0$. Fortunately, the fourth one is a real positive number, $b \approx 6.787$. 

Similarly to the previous section, to study the time evolution of the Borel resummed correlation functions, it is convenient to switch to the rescaled variables~\eqref{resphi}. In terms of these variables, expression~\eqref{twopB} becomes   
\begin{equation}
\left\langle\bar\phi^2\right\rangle_{\rm S}(\bar N) = {\sqrt{\lambda}\over H^2}\int_{0}^{\infty}dz z^be^{-z}\left\langle \phi^2\right\rangle_{\rm B}\left(\frac{z\bar N}{\sqrt{\lambda}H}\right),
\label{twoprB}
\end{equation}
and, correspondingly,
\begin{equation}
\left\langle\bar\phi^4\right\rangle_{\rm S}(\bar N) = {{\lambda}\over H^4}\int_{0}^{\infty}dz z^be^{-z}\left\langle \phi^4\right\rangle_{\rm B}\left(\frac{z\bar N}{\sqrt{\lambda}H}\right).
\label{fourprB}
\end{equation}
As is clear from~\eqref{autonB2} and~\eqref{autonB6}, functions $\left\langle\bar\phi^2\right\rangle_{\rm S}(\bar N)$ and $\left\langle\bar\phi^4\right\rangle_{\rm S}(\bar N)$ do not explicitly depend on the values of the Hubble and coupling constants.    

\begin{figure}[htbp]
    \centering
    \includegraphics[scale=0.52]{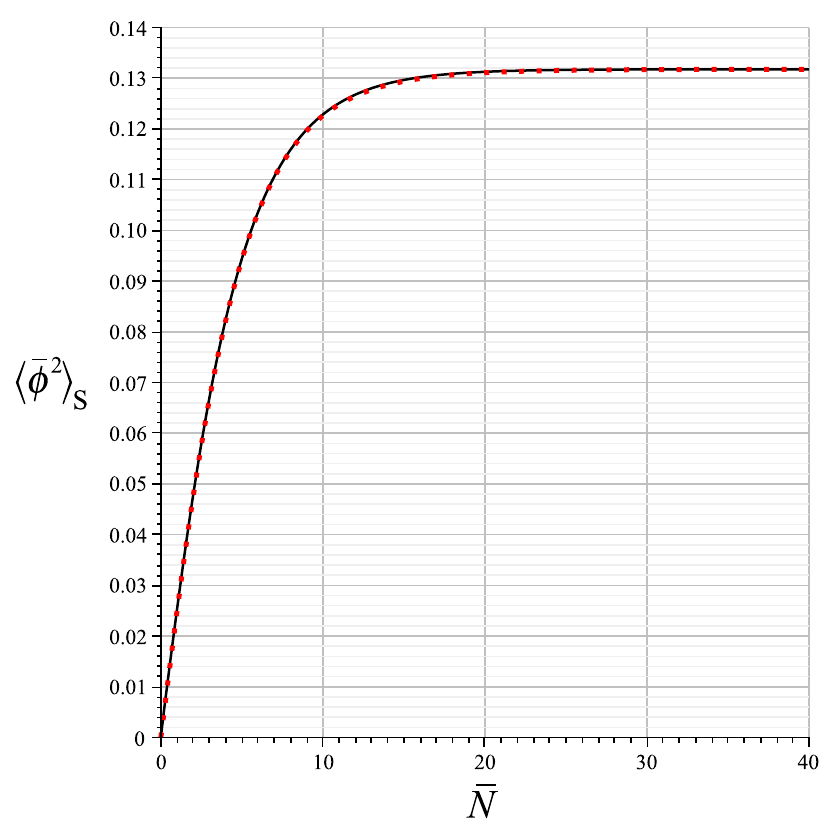}\hfill
    \includegraphics[scale=0.52]{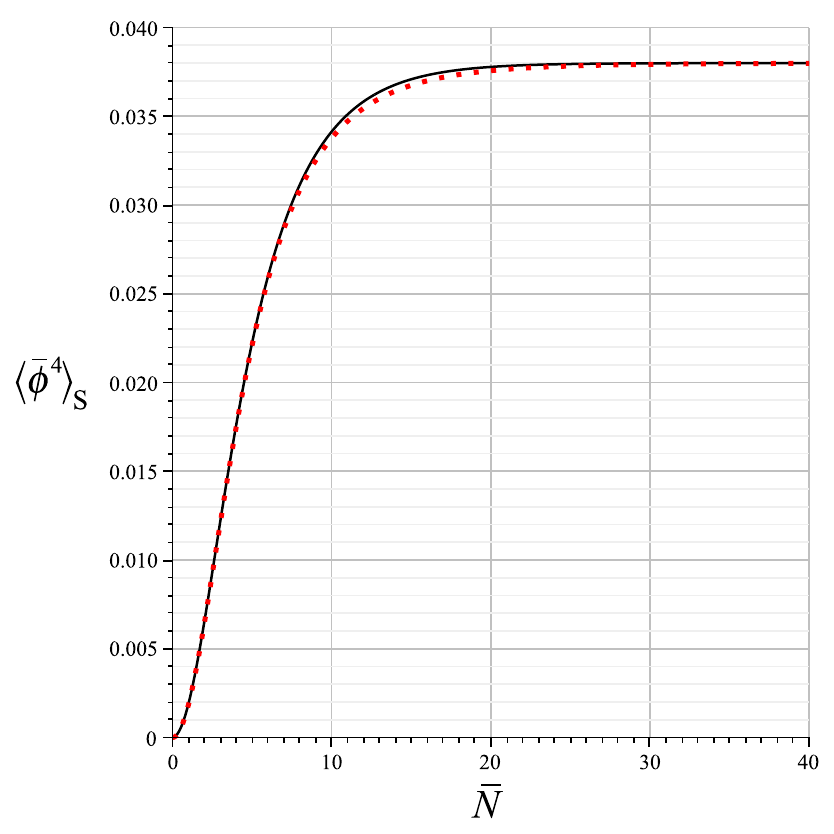}
    \caption{Left: the red dotted line is the plot of the Borel resummation~\eqref{twoprB}, in which $\left\langle \phi^2\right\rangle_{\rm B}$ is the solution~\eqref{autonB2} of the autonomous equation with $b=13.55$, and the black solid line corresponds to $\langle\bar\varphi^2\rangle(\bar N)$, which is calculated from the numerical solution ${\cal P}(\bar\varphi,\bar N)$ of the Fokker--Planck  equation \eqref{FPres}. Right: the red dotted line is the plot of the Borel resummation~\eqref{fourprB}, in which $\left\langle \phi^4\right\rangle_{\rm B}$ is the solution~\eqref{autonB6} with $b=6.787$, and the black solid line corresponds to $\langle\bar\varphi^4\rangle(\bar N)$, which is calculated from the numerical solution ${\cal P}(\bar\varphi,\bar N)$ of the Fokker--Planck  equation \eqref{FPres}.}
    \label{fig:Borel_both}
\end{figure}

Figure \ref{fig:Borel_both} shows the Borel resummed correlation functions obtained by  performing numerical integration in~\eqref{twoprB} and~\eqref{fourprB} with $b=13.55$ and $b=6.787$, respectively. As we can see, the Borel resummation improved the quantitative accuracy of our results. The improvement looks especially good for the four-point function (compare the right panels of Figures~\ref{fig:both} and \ref{fig:Borel_both}): the biggest relative difference between $\left\langle\bar\phi^4\right\rangle_{\rm S}(\bar N)$ and $\left\langle\bar \varphi^4\right\rangle(\bar N)$ occurs around $\bar N\approx11$ and equals $\approx 1.1\%$, as opposed to $9\%$ that we had in the previous section. The two-point function is also reproduced better: the biggest relative difference between $\left\langle\bar\phi^2\right\rangle_{\rm S}(\bar N)$ and $\left\langle\bar \varphi^2\right\rangle(\bar N)$ is only $\approx 0.2\%$. 
%\begin{figure}
%    \begin{center}
%        \includegraphics[scale=0.6]{2pointBorel.pdf}
%        \caption{Comparison of 2-point correlators: the red dotted line is the plot of the Borel resummation~\eqref{twoprB}, in which $\left\langle \phi^2\right\rangle_{\rm B}$ is solution \eqref{autonB2} of the autonomous equation and $b=13.55$, and the black solid line corresponds to $\langle\bar\varphi^2\rangle(\bar N)$, which is calculated from the numerical solution ${\cal P}(\bar\varphi,\bar N)$ of the Fokker-Planck  equation \eqref{FPres}.}\label{fig:three}
%    \end{center}
%\end{figure}
%\begin{figure}
%    \begin{center}
%        \includegraphics[scale=0.6]{4pointBorel.pdf}
%        \caption{Comparison of 4-point correlators: the red dotted line is the plot of the Borel resummation  \eqref{fourprB}, in which $\left\langle \phi^4\right\rangle_{\rm B}$ is solution \eqref{autonB6} of the autonomous equation and $b=6.787$, and the black solid line corresponds to $\langle\bar\varphi^4\rangle(\bar N)$, which is calculated from the numerical solution ${\cal P}(\bar\varphi,\bar N)$ of the Fokker-Planck  equation \eqref{FPres}.}\label{fig:four}
%    \end{center}
%\end{figure}  

\section{Singularities of Borel transforms}\label{singularities}
In this section we consider a slightly different type of Borel transform for the perturbative series~\eqref{sec5} of the four-point function. Instead of $t$, we regard $t^2$, or rather~$N^2$, as an expansion parameter. In this case, the first three terms in the Borel--Le Roy transform of \eqref{sec5} are      
\begin{equation}%\begin{eqnarray}
\left\langle \phi^4\right\rangle_{\rm B}(z) = {3H^4z\over 16\pi^4\Gamma(b+2)}-{3\lambda H^4z^2\over 32\pi^6\Gamma(b+3)} +{53\lambda^2H^4z^3\over 960\pi^8\Gamma(b+4)}+ {\cal O}\bigl(\lambda^3\bigr),
\label{Borel}
\end{equation}%\end{eqnarray}
and the Borel resummed four-point correlator is obtained as
\begin{equation}
{\left\langle\phi^4\right\rangle}_{{\rm S}(N^2)}(N) = \int_0^{\infty}dzz^b e^{-z}\left\langle \phi^4\right\rangle_{\rm B}(zN^2),
\label{standart2}
\end{equation}
where the ${{\rm S}(N^2)}$ subscript indicates that the Borel transform in the integral of \eqref{standart2} was performed with $N^2$ being the expansion parameter.

\begin{figure}
    \begin{center}
        \includegraphics[scale=0.6]{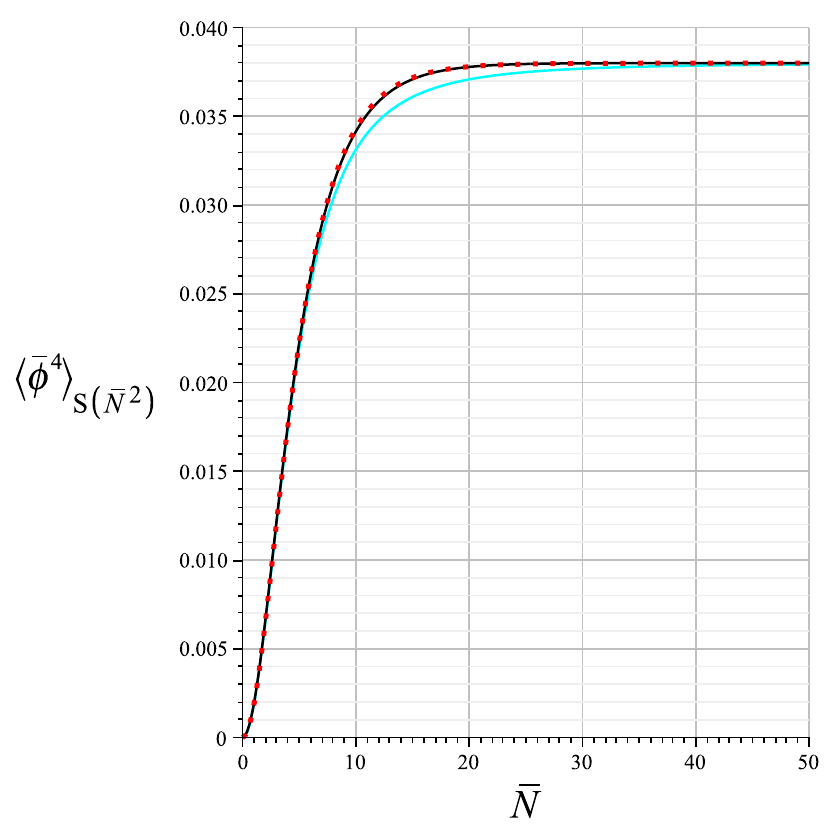}
        \caption{Comparison of four-point correlation functions: the red dotted line and the blue solid line are the plots of the Borel resummation~\eqref{gen1}, in which $\left\langle \phi^4\right\rangle_{\rm B}$ is the solution~\eqref{solution} of the corresponding autonomous equation with $b=2.897$ and  $b=0.4104$, respectively, and the black solid line corresponds to $\langle\bar\varphi^4\rangle(\bar N)$, which is calculated from the numerical solution ${\cal P}(\bar\varphi,\bar N)$ of the Fokker{--}Planck equation \eqref{FPres}.}\label{fig:five}
    \end{center}
\end{figure}

The autonomous equation that reproduces the terms in~\eqref{Borel} has the following form,
%\begin{widetext}
\begin{equation} 
\frac{d\left\langle \phi^4\right\rangle_{\rm B}}{dz}= {3H^4\over 16\pi^4\Gamma(b+2)}-\lambda{\left\langle \phi^4\right\rangle_{\rm B}\over (b+2)\pi^2}+\lambda^2{4(16+23b)\Gamma(b+2)\over 45(b+2)^2(b+3)H^4}\left\langle \phi^4\right\rangle_{\rm B}^2.
\label{Borel1}
\end{equation}
Unlike equations~\eqref{auton5} and \eqref{auton14}, the solution of equation~\eqref{Borel1} can be found explicitly; with the initial condition $\left\langle \phi^4\right\rangle_{\rm B}(0)=0$, we have
%\begin{equation}
%\small{
\begin{align}\label{solution}
\left\langle \phi^4\right\rangle_{\rm B}(z) =& \frac{45H^4}{8\pi^2\lambda} {(b+2)(b+3)\over(16+23b)\Gamma(b+2)}  \\ \nonumber 
& \times\left(1-\sqrt{{(29-8b)\over15(b+3)}}\,{\exp(\sqrt{\dfrac{(29-8b)}{15(b+3)}}\dfrac{\lambda z}{(b+2)\pi^2})+\dfrac{\sqrt{15(b+3)}-\sqrt{29-8b}}{\sqrt{15(b+3)}+\sqrt{29-8b}}\over\exp(\sqrt{\dfrac{(29-8b)}{15(b+3)}}\dfrac{\lambda z}{(b+2)\pi^2})
-\dfrac{\sqrt{15(b+3)}-\sqrt{29-8b}}{\sqrt{15(b+3)}+\sqrt{29-8b}}}\right).
\end{align} %} %\end{equation}
%\end{widetext}\normalsize \noindent 
The parameter $b$ can be fixed by following the same procedure as in the previous section, but here we obtain two values for which the asymptotic value of the Borel summed four-point correlation function~\eqref{standart2} matches with its counterpart in the stochastic picture: $b\approx2.897$ and  \mbox{$b\approx0.4104$}. Figure~\ref{fig:five} shows the rescaled version of the Borel resummation~\eqref{standart2},
\begin{equation}
{\left\langle\bar\phi^4\right\rangle}_{{\rm S}(\bar N^2)}(\bar N) = {{\lambda}\over H^4}\int_0^{\infty}dzz^b e^{-z}\left\langle \phi^4\right\rangle_{\rm B}\left(\frac{z\bar N^2}{\lambda}\right).
\label{gen1}
\end{equation}
As we can see, the correlation function is reproduced with even better accuracy than before: with \mbox{$b=2.897$} the biggest relative difference between ${\left\langle\bar\phi^4\right\rangle}_{{\rm S}(\bar N^2)}$ and $\left\langle\bar \varphi^4\right\rangle$ is only \mbox{$\approx0.6\%$}.

Our main motivation for considering the Borel transform with $N^2$ being the expansion parameter is to compare the singularities of function~\eqref{solution} with the singularities obtained in paper~\cite{Borel-sec}. As mentioned in the previous section, the authors of~\cite{Borel-sec} use high-order Pad\'e approximants for the Borel transforms of the two-point and four-point functions, and, to perform the Borel transform, $N^2$ is used as the expansion parameter. These approximants have alternating poles and zeros along the real negative axis of the complex $z$-plane, suggesting that they inherit the branch cut of the {\it exact} Borel transformation~\cite{Yamada}, which starts at $\lambda z \approx -80$ as indicated by the pole closest to the origin~\cite{Borel-sec}. 

Let us note that in \cite{Borel-sec}, the Borel transform is defined a little differently from what we have been using throughout this paper. For a series that has no $n=0$ term, 
%\begin{equation}
$f(t) = \sum\limits_{n=1}^{\infty}a_n t^n$, 
%\label{5series} \end{equation}
the Borel transform in \cite{Borel-sec} is defined as
\begin{equation}
g_{\rm B}(z)\equiv \sum_{n=1}^{\infty}\frac{a_n}{\Gamma(n)} \, z^{n-1}.
\label{5seriesBJ}
\end{equation}
Hence, it is related to the Borel{--}Le Roy transform~\eqref{series4} in the following way:  
\begin{equation}
f^{(-1)}_{\rm B}(z) \, = z\,g_{\rm B}(z),
\label{BJrel}
\end{equation}
where the superscript indicates the value of the parameter $b$.

It is well known that the position of the singularity of the Borel{--}Le Roy transform is independent of the parameter $b$. This can be seen by noting that for $b_2>b_1$, two Borel{--}Le Roy transforms are related by the integral transformation,   
\begin{equation}
f^{(b_2)}_{\rm B}(z)=\frac{1}{\Gamma(b_2-b_1)}\int_0^1 du(1-u)^{b_2-b_1-1}u^{b_1}f^{(b_1)}_{\rm B}(zu),
\label{as6}
\end{equation}
which can be derived by using the integral definition of the beta function and its relation to the gamma function, \mbox{$B(a,b) = \cfrac{\Gamma(a)\Gamma(b)}{\Gamma(a+b)}$}. Combining~\eqref{BJrel} and \eqref{as6} allows us to express the Borel{--}Le Roy transform \eqref{series4} in terms of the transform~\eqref{5seriesBJ}: as long as $b>-1$,    
\begin{equation}
f^{(b)}_{\rm B}(z)=\frac{z}{\Gamma(b+1)}\int_0^1 du(1-u)^{b}g_{\rm B}(zu).
\label{as7}
\end{equation}
It follows from \eqref{as6} and \eqref{as7} that even though the position of the singularity does not change, its type might not stay the same: for example, the order of a pole can change, a pole can turn into a branch point or vice versa, and so on. 

Let us now look at the singularities of function \eqref{solution} with $b=2.897$. Since it is not an exact Borel transform, but an approximation derived from the autonomous equation constructed from only the first three terms of its series expansion, we should not expect to see the same singularities that are obtained in \cite{Borel-sec} from the high-order Pad\'e approximants. Indeed, \eqref{solution}  has no branch point, and the only singularities are simple poles located at \mbox{$\lambda z\approx-98.87+ i\,1183\,k$}, where $k$ is an integer. The poles with $k\neq 0$ are most likely an artifact of our approximation. On the other hand, the pole at $-98.87$ might correspond to the branch point at $-80$ found in \cite{Borel-sec}, being shifted due to \eqref{solution} not being an exact Borel transform. 
 
\section{The truncated Schwinger--Dyson-type equations}\label{S_D_eqs}
In the concluding section of this paper, we propose an alternative method for extracting the perturbative coefficients of correlation functions in de Sitter space. These coefficients constitute the starting point for implementing the autonomous equation and associated resummation techniques.

As we mentioned in the introduction, an infinite system of the Schwinger--Dyson equations for a massive self-interacting scalar field, $m^2\phi^2/2+\lambda\phi^4/4$, in the Euclidean zero-dimensional field theory~\cite{zero} was considered in~\cite{Schwinger-D,Bender:1988bp,Okopinska:1990pt}. The even-point correlation functions satisfy an infinite system of algebraic Schwinger-Dyson equations, which has the form of the recursion relation
\begin{equation}
\lambda \bigl\langle \phi^{2n+2}\bigr\rangle=-m^2 \bigl\langle \phi^{2n} \bigr\rangle+(2n-1) \bigl\langle \phi^{2n-2}\bigr\rangle.
\label{S-D2}
\end{equation}
By truncating this infinite tower of equations, the authors of~\cite{Schwinger-D,Bender:1988bp} found a solution by rational approximants, the so-called Schwinger--Dyson approximants. Note that the exact expressions for non-perturbative correlation functions $\bigl\langle \phi^{2n}\bigr\rangle$ are known and can be expressed in terms of the modified Bessel functions of the second kind. This makes it possible to study the convergence properties of the introduced Schwinger{--}Dyson approximants. On the other hand, the exact expressions for $\bigl\langle \phi^{2n}\bigr\rangle$ have the same form as the asymptotic values of the corresponding correlation functions obtained within the Starobinsky--Yokoyama stochastic approach; see~\cite{we1,we2,Beneke}. This motivates studying a system analogous to~\eqref{S-D2}, which involves the time-dependence of correlation functions arising in the stochastic picture.

The Schwinger{--}Dyson-type equations are derived from the Fokker{--}Planck equation~\eqref{F-P} expressed in terms of moments, which takes the following form~\cite{Tsam-Wood, Q-S, we3}:
\begin{equation}
\frac{\partial}{\partial t} \, \langle \varphi^{2n} \rangle = -\frac{2nm^2}{3H}\langle \varphi^{2n} \rangle-\frac{2n\lambda}{3H}\langle \varphi^{2n+2} \rangle + n(2n-1)\frac{H^3}{4\pi^2}\langle \varphi^{2n-2} \rangle.
\label{diff_first-order_FPK}
\end{equation}
Denoting $f_n(t)\equiv\langle\varphi^{2n}\rangle$(t), we have 
\begin{equation}\label{eq_for_f_n}
\,\,\, \dot{f}_n = - \alpha n f_n - \beta nf_{n+1} + \gamma n ( 2n - 1 ) f_{n-1}, \end{equation}
where
\begin{equation}\label{abc_notation} 
\alpha \equiv\frac{2m^2}{3H}, \qquad \beta \equiv\frac{2\lambda}{3H}, \qquad \text{and} \qquad \gamma \equiv\frac{H^3}{4\pi^2}.
\end{equation} 
This system of equations is infinite, but for simplicity one can truncate it by setting the highest moment to zero, $f_n(t) = 0$, in the last kept equation. We propose a method to solve the truncated tower of equations such that the perturbative expansion is correct up to a specified order in $\lambda$. The idea is to combine the finite set of equations from the truncated system into a single differential equation, solve it, and express all retained moments in terms of that solution. We illustrate our method with simple examples.

For instance, in the massless case, $\alpha =0$, our system of equations~\eqref{eq_for_f_n} truncated at the first two equations takes a simple form:
\begin{equation}\label{2tr_f_1_f_2}
    \dot{f}_1 = - \beta\, f_2 + \gamma\, f_0, \qquad  \dot{f}_2 = 6\gamma\, f_1. 
\end{equation} We impose $\left. f_n (t) \right|_{n\geq 3} = 0$, and $f_0 =1$ is the normalization condition. Using the second equation to eliminate $f_1(t)$ and substituting into the first equation yields
\begin{equation}
\ddot{f}_2 + 6 \beta \gamma \, f_2 - 6 \gamma^2=0.
\end{equation} The solution to this trivial equation is $f_2(t) = c_1 \e^{\sigma_1 t} + c_2 \e^{\sigma_2 t} + \dfrac{\gamma}{\beta}$. The unknown $\sigma_i$ in exponents are purely imaginary since they satisfy $\sigma_i^2 + 6\, \beta \gamma =0$. Thus, the solution takes the following form:
\begin{equation}\label{2tr_sol_f_2}
f_2(t) = c_1 \, \sin{( \omega t)} + c_2 \, \cos{(\omega t)} + \frac{\gamma}{\beta}\, , %\quad \omega^2 := 6\, bc,
\end{equation} where we denote $\omega^2 = 6\, \beta\gamma$, and from~\eqref{2tr_f_1_f_2} we obtain
\begin{equation}\label{2tr_sol_f_1}
f_1(t) = \frac{1}{6\gamma} \, \dot{f}_2 = \frac{c_1 \, \omega}{6\gamma} \cos{(\omega t)} - \frac{c_2 \, \omega}{6\gamma} \sin{(\omega t)}.
\end{equation} To define coefficients $c_i$, we use the initial conditions $\left. f_n(t)\right|_{t=0} =0$, resulting in $c_1=0$ and $ c_2 = - \dfrac{\gamma}{\beta}$. Thus, 
\begin{equation}\label{2tr_sol_f_1_f_2_full}
f_1(t) =\frac{\omega}{6\beta} \sin{\bigl(\omega t\bigr)} \qquad \text{and} \qquad f_2(t) = -\frac{\gamma}{\beta} \, \cos{\bigl(\omega t\bigr)} + \frac{\gamma}{\beta}\,. %, \qquad \omega^2 := 6\, bc.
\end{equation} Expanding these in the small $t$ regime and substituting $\beta$ and $\gamma$ from~\eqref{abc_notation}, we have 
\begin{align}\label{two_truncated_massless_series_f_1}%\begin{equation}
f_1(t) &\approx \frac{\omega^2}{6\, \beta} \, t - \frac{\omega^4}{36 \, \beta} \, t^3 \colorgapunderline{red}{+ \frac{\omega^6}{720\, \beta }}\, t^5 + {\cal O} \bigl(t^7\bigr) = \gamma \, t - \beta \gamma^2 \,t^3 \colorgapunderline{red}{+ \frac{3}{10} \, \beta^2 \gamma^3}\, t^5 + {\cal O} \bigl(t^7\bigr) \\ 
& \stackrel{\eqref{abc_notation}}{=} \, \frac{H^3t}{4\pi^2} - \frac{\lambda H^5 t^3}{24 \, \pi^4} \colorgapunderline{red}{+ \, \frac{\lambda^2 H^7 t^5}{480 \, \pi^6}} +  {\cal O} \bigl(t^7\bigr), \nonumber \\
\label{two_truncated_massless_series_f_2}
f_2(t) & \approx \frac{\gamma\, \omega^2}{2\beta} \, t^2 \colorgapunderline{red}{- \frac{\gamma\, \omega^4}{24 \, \beta}} \, t^4 + {\cal O} \bigl(t^6\bigr) %\\ \nonumber & 
= 3\gamma^2 \, t^2  \colorgapunderline{red}{- \frac{3}{2} \, \beta \gamma^3} \, t^4  + {\cal O} \bigl(t^6\bigr)\nonumber\\ 
& \stackrel{\eqref{abc_notation}}{=} \, \frac{3 H^6 t^2}{16 \, \pi^4} \colorgapunderline{red}{- \frac{ \,\lambda H^8 t^4}{64 \, \pi^6}} +  {\cal O} \bigl(t^6\bigr).
\end{align}%\end{equation} 
%Substituting $\beta$ and $\gamma$ from~\eqref{abc_notation}, we have 
%\begin{equation} \begin{align} f_1(t) & \approx \, \frac{H^3t}{4\pi^2} - \frac{\lambda H^5 t^3}{24 \, \pi^4} \colorgapunderline{red}{+ \, \frac{\lambda^2 H^7 t^5}{480 \, \pi^6}} +  O \bigl(t^7\bigr), \\  f_2(t) \approx \frac{3 H^6 t^2}{16 \, \pi^4} \colorgapunderline{red}{- \frac{ \,\lambda H^8 t^4}{64 \, \pi^6}} +  O \bigl(t^6\bigr).\end{align}%\end{equation}
The obtained expansion for $f_1(t)$ matches the result within perturbative quantum field theory computations \cite{Tsam-Wood,we,Onemli,Karakaya-Onemli,we2,we3,Kahya} up to the linear order in $\lambda$, while $f_2(t)$ matches it only up to the first term. They differ starting from the next orders in $\lambda$, which we have underlined in red. This is not surprising, since we left out $f_3(t)$ in the second equation~\eqref{2tr_f_1_f_2}. The omitted term would enter multiplied by $\lambda$, so its absence affects higher-order contributions. Given this, our expanded $f_1(t)$ is accurate only up to linear order in $\lambda$.

To correctly capture the next perturbative orders, one must truncate our system~\eqref{eq_for_f_n} at the third equation. Our strategy is the same. We combine all equations within the truncated tower into a single differential equation for the kept highest moment, which is $\dddot{f_3} + 36\, \beta\gamma \, \dot{f_3} - 90\, \gamma^3 =0$. The solution to the corresponding homogeneous equation has the form $f^{\text{hom}}_3(t)= c_1 \e^{\sigma_1 t} + c_2 \e^{\sigma_2 t} + c_3 \e^{\sigma_3 t}$, and  $\sigma_i$ satisfy $\sigma_i^3 + 36 \beta \gamma\, \sigma_i = 0$. This equation has three roots: $\sigma_i = 0$ and $\sigma_i^2 = -36 \beta \gamma$. A particular solution to the inhomogeneous equation is $f^{\text{inhom}}_3(t)=\dfrac{5\gamma^2t}{2\beta}$, and the full solution is $f_3(t) = c_1 + c_2 \sin{(\omega t)} + c_3 \cos{(\omega t)} + \dfrac{5\gamma^2t}{2\beta}$ with $\omega^2=36 \beta \gamma$. Since one of the roots is zero, another way to solve it is to reduce the order of the equation. By setting $\dot{f}_3 = y(t)$, one obtains a corresponding equation that can be solved straightforwardly. 
%%%%%$\ddot{y} + 36 bc \, y - 90c^3 = 0$ and seeks a solution of the form $y(t) = \tilde{c}_1 e^{\alpha_1 t} + \tilde{c}_2 e^{\alpha_2 t} + 5c^2/2b$ The unknown $\alpha_i$ are imaginary since they satisfy $\alpha_i^2 + 36\, bc =0$, and the solution takes the form $y(t)=c_1 \, \sin{( \omega t)} + c_2 \, \cos{(\omega t)} + 5c^2/2b$ with $\omega^2=36 bc$. 
Eventually, with the chosen initial conditions, $\left. f_n(t)\right|_{t=0} =0$, we find $c_1 = c_3 = 0$, $c_2 = - \dfrac{5\gamma^2}{2\beta\omega}$, and
\begin{align}\label{3tr_f_1_final}
f_1(t) & = \frac{5\gamma}{6}\, t  + \frac{\gamma}{6\, \omega}\, \sin{(\omega t)}, \\ \label{3tr_f_2_final}
%\qquad% 
f_2(t) & = \frac{\gamma}{6\beta} - \frac{\gamma}{6\beta} \,  \cos{(\omega t)}, \\ 
\label{3tr_f_3_final}
%\qquad% 
f_3(t) & = \frac{5 \gamma^2}{2\beta} \, t - \frac{5\gamma^2}{2\beta\,\omega} \, \sin{(\omega t)}.
\end{align} %\end{equation}
By a trivial check, expanding these expressions for small $t$ establishes agreement with our previous results~\cite{we2,we3} for $f_1(t)$ up to $\lambda^2$, $f_2(t)$ up to $\lambda$, and $f_3(t)$ up to $\lambda^0$ orders.

We provide the last detailed derivation for the truncated system of four equations. Once again, in the last retained equation we set the highest moment to zero and combine equations into a single differential equation:
%\begin{equation}
$\ddddot{f_4} + 120 \beta \gamma \ddot{f}_4 + 504 \beta^2 \gamma^2 f_4 - 2520 \gamma^4=0$.  
%\end{equation} 
Its solution is $ f_4(t) = \sum\limits_{i=1}^4 c_i \e^{\sigma_i t} + \dfrac{5\gamma^2}{\beta^2}$, where $\sigma_i$ satisfy the following biquadratic equation, $\sigma_i^4 + 120 \beta \gamma \, \sigma_i^2 + 504 \beta^2 \gamma^2 =0 $, all of which are purely imaginary. Therefore, it is more convenient to write the solution in the trigonometric form: 
%\begin{equation}
%\begin{align}\label{equation_for_f_4_omega}
$f_4(t) =  c_1 \sin{(\omega_1 t)} + c_2 \cos{(\omega_1 t)} %\\ \nonumber & \,\,\, 
+ c_3 \sin{(\omega_2 t)} + c_4 \cos{(\omega_2 t)} + \dfrac{5\gamma^2}{\beta^2}$, 
%\end{align}%\end{equation} 
and $\omega_1$ and $\omega_2$ satisfy the following relations: %$\omega_i^4-120\, bc \, \omega_i^2 + 504 \, b^2c^2=0$
\begin{equation}\label{omega_equation_1}
\omega_1^2 + \omega_2^2 = 120 \beta \gamma  \qquad \text{and} \qquad (\omega_1 \omega_2)^2 = 504 \beta^2 \gamma^2 .   
\end{equation} 
Using the initial condition $f_n(0) = 0$ and solving the resulting algebraic system, we obtain $c_1 = c_3 =0$, $c_2 = \dfrac{5\gamma^2\, \omega_2^2}{\beta^2\, \bigl(\omega_1^2-\omega_2^2\bigr)}$, and $c_4 = - \, \dfrac{5\gamma^2\, \omega_1^2}{\beta^2\, \bigl(\omega_1^2-\omega_2^2\bigr)}$. 
%\begin{equation}\begin{gathered} c_1 = c_3 =0, %\quad \text{and} \quad
%\qquad c_2 = \frac{5c^2\, \omega_2^2}{b^2\, \bigl(\omega_1^2-\omega_2^2\bigr)}, \\ %\quad \text{with} \quad 
%c_4 = - \, \frac{5c^2\, \omega_1^2}{b^2\, \bigl(\omega_1^2-\omega_2^2\bigr)}.
%\end{gathered} \end{equation} 
Therefore, the resulting expression for the solution to our truncated system is
%\begin{widetext} 
\begin{align}\label{f_1_full_answer}
f_1(t) & = \frac{\omega_1 \omega^2_2 \, \bigl( \omega_1^2 - 114\, \beta \gamma \bigr) }{504\,\beta^2\gamma\, \bigl(\omega_1^2-\omega_2^2\bigr)} \, \sin{(\omega_1 t)} - \frac{\omega^2_1 \omega_2 \, \bigl( \omega_2^2 - 114\, \beta \gamma \bigr)}{504\,\beta^2 \gamma\, \bigl(\omega_1^2-\omega_2^2\bigr)} \, \sin{(\omega_2 t )}\,, \\
\label{f_2_full_answer}
f_2(t) & = - \frac{\,\omega_2^2 \, \bigl( \omega_1^2 - 84\, \beta \gamma \bigr)}{84\,\beta^2\, \bigl(\omega_1^2-\omega_2^2\bigr)} \, \cos{(\omega_1 t)} + \frac{\,\omega^2_1 \, \bigl( \omega_2^2 - 84 \, \beta \gamma \bigr) }{84\, \beta^2\, \bigl(\omega_1^2-\omega_2^2\bigr)} \, \cos{(\omega_2 t)}\, + \frac{\gamma}{\beta}, \\
\label{f_3_full_answer}
f_3(t) & = - \frac{5 \gamma\, \omega_1 \omega^2_2}{28 \, \beta^2\, \bigl(\omega_1^2-\omega_2^2\bigr)} \, \sin{(\omega_1 t)} + \frac{5\gamma \, \omega^2_1 \omega_2}{28\, \beta^2\, \bigl(\omega_1^2-\omega_2^2\bigr)} \, \sin{(\omega_2 t)}\,, \\
\label{f_4_full_answer}
f_4(t) & =  \frac{5\gamma^2\, \omega_2^2}{\beta^2\, \bigl(\omega_1^2-\omega_2^2\bigr)} \, \cos{(\omega_1 t )} - \, \frac{5\gamma^2\, \omega_1^2}{\beta^2\, \bigl(\omega_1^2-\omega_2^2\bigr)} \, \cos{(\omega_2 t)}\, + \frac{5\gamma^2}{\beta^2}.
\end{align} 
Expanding these expressions as power series in $t$, we obtain
\begin{align}\label{f_1_full_expanded}
    & f_1 (t) \approx \frac{\omega_1^2 \omega_2^2}{504\, \beta^2 \gamma} \, t - \frac{\omega_1^2 \omega_2^2}{3024\, \beta^2 \gamma} \, \bigl(\omega_1^2 + \omega_2^2 - 114\, \beta \gamma \bigr) \, t^3 \\ \nonumber 
    & \qquad \quad + \frac{\omega_1^2 \omega_2^2}{60480\, \beta^2 \gamma} \Bigl( \bigl( \omega_1^2 + \omega_2^2\bigr)^2 - \omega^2_1\omega_2^2 - 114 \, \beta \gamma \, \bigl(\omega_1^2 +\omega_2^2\bigr) \Bigr) t^5 \\ \nonumber 
    & %\qquad \quad
    - \frac{\omega_1^2 \omega_2^2}{2540160 \beta^2\gamma} \biggl( \bigl( \omega_1^2 + \omega_2^2\bigr) \Bigl( \bigl( \omega_1^2 + \omega_2^2\bigr)^2 - 2 \omega^2_1\omega_2^2 \Bigr) - 114 \beta \gamma \Bigl( \bigl(\omega_1^2 +\omega_2^2\bigr)^2 - \omega_1^2\omega_2^2 \Bigr) \biggr) t^7 + {\cal O} \bigl(t^{9}\bigr), \\
    & f_2 (t) \approx \frac{\omega_1^2 \omega_2^2}{168\, \beta^2} \, t^2 - \frac{\omega_1^2 \omega_2^2}{2016\, \beta^2} \, \bigl(\omega_1^2 + \omega_2^2 - 84\, \beta \gamma \bigr) \, t^4 \\ \nonumber 
    & \qquad \quad 
    + \frac{\omega_1^2 \omega_2^2}{60480\, \beta^2} \Bigl( \bigl(\omega_1^2 + \omega_2^2\bigr)^2 - \omega_1^2 \omega_2^2 - 84\, \beta \gamma \, \bigl(\omega_1^2 + \omega_2^2\bigr) \Bigr) \, t^6  + {\cal O} \bigl(t^{8}\bigr), \\
    \label{f_3_full_expanded}
    & f_3 (t) \approx \frac{5 \gamma \, \omega_1^2 \omega_2^2}{168\, \beta^2}\, t^3 - \frac{\gamma\, \omega_1^2 \omega_2^2}{672\, \beta^2} \bigl(\omega_1^2 + \omega_2^2\bigr) \, t^5 + O\bigl(t^{7}\bigr), \\
    %\end{align} \begin{align}
    \label{f_4_full_expanded} 
    & f_4 (t) \approx \frac{5 \gamma^2 \omega_1^2\omega_2^2}{24\, \beta^2}\, t^4 + {\cal O}\bigl(t^{6}\bigr).
\end{align} %\end{widetext}
Note that only the combinations from~\eqref{omega_equation_1} appear in the expansions above, so we do not need to determine $\omega_i$ explicitly. Thus, the final expression for our solution in the small $t$ regime appears to be
\begin{align}\label{f_1_answer_expanded}
    f_1(t) & \approx \, \gamma\, t - \beta \gamma^2 t^3 + \frac{9}{5} \, \beta^2 \gamma^3 \, t^5 - \frac{159}{35} \, \beta^3 \gamma^4\, t^7 + {\cal O} \bigl(t^{9}\bigr) \\ \nonumber 
    & \stackrel{\eqref{abc_notation}}{=} \, \frac{H^3t}{4\pi^2} - \frac{\lambda H^5 t^3}{24 \, \pi^4} + \frac{\lambda^2 H^7 t^5}{80 \, \pi^6} - \frac{53 \lambda^3 H^9 t^7}{10080 \, \pi^8} + {\cal O} \bigl(t^{9}\bigr), \\
    \label{f_2_answer_expanded}
    f_2(t) & \approx  \, 3 \gamma^2\, t^2 - 9 \beta \gamma^3 \, t^4 + \frac{159}{5} \, \beta^2 \gamma^4 \, t^6 + {\cal O}\bigl(t^{8}\bigr) \\ \nonumber 
    & \stackrel{\eqref{abc_notation}}{=} \frac{3 H^6 t^2}{16 \, \pi^4} - \frac{3 \,\lambda H^8 t^4}{32 \, \pi^6} +  \frac{53 \, \lambda^2 H^{10} \, t^6}{960 \, \pi^8} + {\cal O} \bigl(t^8\bigr), \\
    \label{f_3_answer_expanded}
    f_3(t) & \approx \, 15 \gamma^3\, t^3 - 90 \beta \gamma^4\, t^5 + {\cal O} \bigl(t^{7}\bigr) \stackrel{\eqref{abc_notation}}{=} \, \frac{15 H^9 t^3}{64 \, \pi^6} - \frac{15 \, \lambda H^{11} t^5}{64 \, \pi^8} + {\cal O} \bigl(t^{7}\bigr), \\
    \label{f_4_answer_expanded}
    f_4(t) & \approx \, 105 \gamma^4 \, t^4 + {\cal O} \bigl(t^{6}\bigr)
    \stackrel{\eqref{abc_notation}}{=} \, \frac{105 H^{12} t^4}{256 \, \pi^8} + {\cal O} \bigl(t^{6}\bigr).
\end{align} %\end{widetext} 
The presented orders coincide with our previous results in~\cite{we2,we3} and agree with~\eqref{sec2} and~\eqref{sec6} up to $\lambda^2${-}order when expanded in the small self-interaction coupling constant regime. Let us note that at the three-loop level, the coefficient $- 43/10080$ extracted from the expansion of our solution~\eqref{sec2} to the autonomous equation for $\bigl\langle \phi^2 (t,\x\,) \bigr\rangle$ is quite close to the corresponding coefficient in~\eqref{f_1_answer_expanded}. If one proceeds with the extension of our results~\eqref{f_1_full_expanded}{--}\eqref{f_4_full_expanded} to the next orders, the obtained contributions will differ from the results gained using the Fokker{--}Planck equation via the routine proposed in~\cite{we3} by approximately a factor of two. This is a quite convenient result for perturbation theory in the small self-interaction coupling constant regime. 

To extract the next orders using our presented method here, one obtains the corresponding single differential equation %with $\left. f_\n (t) \right|_{\forall n\geq 6} = 0$: \begin{equation}
$ \overset{\textbf{.....}}{f_5} + 300\, \beta \gamma \, \dddot{f_5} + 6984\, \beta^2 \gamma^2 \, \dot{f}_5 - 113400\, \gamma^5 = 0$. %\end{equation} 
Analogously to the previous example of the three kept equations in the truncated system, the solution to the homogeneous equation is $f^{\text{hom}}_5(t) = \sum\limits_{i=1}^5 \, c_i\, \e^{\sigma_i t}$, where $\sigma_i$ satisfy $\sigma_i^5 + 300\, \beta \gamma \, \sigma^3_i + 6984\, \beta^2\gamma^2 \, \sigma_i =0$, which has five roots: four are purely imaginary and one is zero. Consequently, the homogeneous solution can be cast in the trigonometric form with $\omega_1^2 + \omega_2^2 = 300 \, \beta \gamma$ and $(\omega_1 \omega_2 )^2 = 6984 \, \beta^2 \gamma^2$, while the inhomogeneous solution is $f^{\text{inhom}}_5(t) = \dfrac{1575\, \gamma^3 t}{97\beta^2}$. The full solution together with the outlined procedure for fixing coefficients using the chosen initial values leads to expressions that, when expanded again, contain only combinations of $\omega_i$ (obviously, a change of variables reduces the equation to a fourth-order derivative equation and yields the same solution). The resulting moments expansions agree with those proposed to extract from the Fokker--Planck equation in~\cite{we3}.

The presented treatment uses only trivial algebraic operations and differentiation. Unlike the Schwinger--Keldysh (or ``in-in") formalism, our method does not require handling a large number of diagrams with combinatorial coefficients, nor numerous integrations involving various Green's functions and vertices. Similarly, it avoids the nested integrals typical of the Yang--Feldman equation. Our massless functions $f_n(t)$ obtained in this way exhibit secular growth. Consequently, they are useful only for computing the first few coefficients in the perturbative series. Nevertheless, the simplicity with which these coefficients can be extracted from the Schwinger--Dyson-type relations~\eqref{eq_for_f_n} is noteworthy.

However, we could treat our system of equations~\eqref{eq_for_f_n} in another way. Instead of setting the highest moment to zero in the last kept equation, %$f_\n=0$ for all $n\geq n_0$, 
let us use the Gaussian approximation for the highest moment $n_0$:
\begin{equation}f_{n_0}(t) = (2n_0-1)!! \, \bigl(f_1(t)\bigr)^{n_0}. \label{Gauss} \end{equation} 
In this case, we again obtain a closed system of first-order differential equations for the functions $f_1,...,f_{n_0-1}$. Let us consider two cases. In the simplest case $n_0=2$, we only have one equation:
\begin{equation}
\dot{f}_1 = -3 \beta \, f_1^2 + \gamma \stackrel{\eqref{abc_notation}}{=} -\frac{2\lambda}{H} \, f_1^2 +\frac{H^3}{4\pi^2}. 
\label{Gauss2}
\end{equation}
This is the equation for $\bigl\langle \phi^2(t,\x\,)\bigr\rangle$ within the Hartree--Fock (or Gaussian) approximation~\cite{Star,Star-Yok}; see~\eqref{auton3}. This approximation is known for the resummation of the so-called cactus-type diagrams; see~\cite{Polina_thesis} for higher orders.

The next case of our interest is $n_0 = 3$. Using the Gaussian form~\eqref{Gauss}, plugging $f_3 = 15 f_1^3$ into the second equation of~\eqref{eq_for_f_n}, and combining with the first one, we obtain the equation $\ddot{f}_1=30\, \beta^2 f_1^3 - 6\, \beta \gamma \, f_1$. Multiplying this equation by $2\dot{f_1}$, we find the first integral, meaning that $\dot{f}_1^2-15 \beta^2 \, f_1^4 + 6 \beta \gamma \, f_1^2 = \text{const}$. With the initial conditions $f_1(0)=0$ and $f_2(0)=0$, it follows from~\eqref{eq_for_f_n}  that $\dot{f}_1(0) = \gamma$. Therefore, the constant in the first integral is determined to be $\gamma^2$, resulting in
\begin{equation}
\dot{f}_1=\sqrt{15\beta^2 \, f_1^4 - 6 \beta \gamma \, f_1^2 + \gamma^2 \,}. \label{c3}
\end{equation}
Now, we would like to derive our autonomous equation that has a polynomial structure and includes terms up to the second order in $\lambda$. By expanding the right-hand side of~\eqref{c3} to second order in $\beta$, we get
%\begin{widetext}
\begin{equation}
\dot{f}_1 \approx \gamma -3 \beta \, f_1^2 + \frac{3\beta^2}{\gamma}\, f_1^4 %+ O \bigl(b^3\bigr) 
\stackrel{\eqref{abc_notation}}{=} \frac{H^3}{4\pi^2} - \frac{2\lambda}{H} \, f_1^2 + \frac{16 \lambda^2 \pi^2}{3 H^5} \, f_1^4.
\label{auton0}
\end{equation}%\end{widetext}
Hence, we recover our autonomous equation~\eqref{auton5}. 

Actually, in order to obtain~\eqref{auton9} we solve the linearized equation~\eqref{auton8}, rather than solving~\eqref{auton5}. We can obtain it directly from~\eqref{c3}. Let us represent the function $f_1(t)$ as $f_1 (t) = f_{\,\text{HF}} (t) +\tilde{f} (t)$, where $f_{\,\text{HF}}(t)$ is the solution to~\eqref{Gauss2} within the Hartree--Fock approximation, and function $\tilde{f}$ is a small correction to it. Substituting this expression into equation~\eqref{c3}, rewritten as $\dot{f}_1^2=15\beta^2 f_1^4-6\beta\gamma f_1^2 + \gamma^2$, using~\eqref{Gauss2}, and keeping only terms up to second order in the parameter $\beta$, we obtain the following linearized first-order differential equation for $\tilde{f}(t)$:
\begin{equation} \dot{\tilde{f}}=-6\beta \, f_{\,\text{HF}}\tilde{f} + \frac{3\beta^2}{\gamma}f_{\,\text{HF}}^4 = -\frac{4\lambda}{H}\, f_{\,\text{HF}}\tilde{f}+\frac{16\pi^2\lambda^2}{3H^5}\, f^4_{\,\text{HF}}. \label{unname3} \end{equation}
With the appropriate changes of notation, one can see that~\eqref{unname3} coincides with~\eqref{auton8}. Therefore, we have obtained an alternative derivation of our autonomous equation~\eqref{auton5} and its linearized version~\eqref{auton8}.

The $n_0 = 3$ case can also be used to derive the autonomous equation for the four-point function. By using the Gaussian approximation for the six-point function, $f_3 = 15 f_1^3$, the second equation of~\eqref{eq_for_f_n} becomes $\dot{f}_2 = - 30\beta f_1^3 +6\gamma f_{1}$, and its derivative yields
\begin{equation} 
\,\,\, \ddot{f}_2 = - 90\beta f_1^2\dot{f}_1 +6\gamma \dot f_{1}.
\label{fourp2}
\end{equation}
Substituting the right side of the first equation of~\eqref{eq_for_f_n} into \eqref{fourp2} and neglecting the term quadratic in $\beta$ gives
\begin{equation} 
\,\,\, \ddot{f}_2 = - 90\beta\gamma f_1^2 -6\beta\gamma f_{2}+6\gamma^2.
\label{fourp3}
\end{equation}
Since at this stage we are interested in deriving the autonomous equation to linear order in $\lambda$, in the first term of the right side of \eqref{fourp3} we can use the zeroth order expression for the four-point function, $f_2 = 3 f_1^2$,  
\begin{equation} 
\,\,\, \ddot{f}_2 = - 36\beta\gamma f_2 +6\gamma^2.
\label{fourp4}
\end{equation}
By multiplying this equation by $2\dot{f_2}$ and integrating, we obtain \mbox{$\dot{f}_2^2 + 36\beta\gamma f_2^2 - 12\gamma^2f_2 = \text{const}$}. With the initial conditions $f_1(0)=0$, $f_2(0)=0$ and $f_3(0)=0$, it follows from~\eqref{eq_for_f_n} that $\dot{f}_2(0) =0$; hence, we should set the integration constant to zero. Therefore,
\begin{equation} 
\,\,\, \dot{f_2} = \sqrt{-36\beta\gamma f_2^2 + 12\gamma^2f_2 \,}.
\label{fourp6}
\end{equation}
Expanding to first order in $\lambda$ gives
\begin{equation} 
\,\,\, \dot{f_2} \approx 2\sqrt{3}\gamma \, f_2^{1/2}-3\sqrt{3}\beta \, f_2^{3/2} \stackrel{\eqref{abc_notation}}{=} \frac{\sqrt{3}H^3}{2\pi^2} f_2^{1/2}-\frac{2\sqrt{3}\lambda}{H} f_2^{3/2}.
\label{fourp6_2}
\end{equation}
It is easy to check that the above equation coincides with equation~\eqref{auton12} with the appropriate coefficients~\eqref{coeff1}.

Various truncation schemes for the Schwinger--Dyson equations have been proposed over the years. For example, in paper \cite{Okopinska:1990pt}, for Euclidean zero-dimensional massive $\phi^4$ theory,  the author considers three sets of Schwinger--Dyson equations: for full correlation functions (expectation values of products of fields), connected correlation functions, and 1PI correlation functions. Each of these sets is truncated by setting the respective highest $n$-point function to zero in the last kept equation. Unlike truncation of full correlation functions, the cases of connected  and 1PI functions give good convergence to the exact values in both weak and strong coupling regimes, with the best convergence obtained in the case of 1PI functions. More recently, Schwinger--Dyson equations for connected correlation functions for various field theories in zero-dimensional Euclidean space were also studied in \cite{Bender:2022eze,Bender:2023ttu}, where the authors used two truncation prescriptions. One of the prescriptions was to set the highest $n$-point function to zero in the last kept equation, and the other one to replace it with its asymptotic behavior for large $n$. It was shown that the second prescription improves the convergence of approximants to a value that is very close to the exact result. In this section we considered two truncation schemes for the infinite system of Schwinger--Dyson-type equations of stochastic inflation. The first one, where the highest $n$-point function is set to zero, reproduces the perturbative series for the correlation functions at early times~($\sqrt\lambda t\ll1$), but does not give the correct behavior at late times ($\sqrt\lambda t\gg1$). In the second scheme, we used the Gaussian approximation for the highest moment instead of setting it to zero. Thus, the second prescription preserves more information about the full theory and, therefore, gives qualitatively correct behavior at both early and late times. The lowest truncation \eqref{Gauss2} already gives a solution that approaches a finite value at late times, and this behavior persists as we consider higher truncations.

\section{Conclusion}\label{conclusion}
Perturbative expressions for the correlation functions of the long-wavelength part of a spectator scalar field on de Sitter background are obtained by different authors using different techniques, and all the results are in agreement; see~\cite{Marolf,Gautier,Hollands,Akhmedov,Garbrecht,Gautier1,Markkanen,Kitamoto,Moreau,Markkanen1}. In~the present paper, by using the truncated system of Schwinger--Dyson-type equations, we demonstrated a very simple way to extract the perturbative coefficients of the equal-time correlation functions for the theory~\eqref{action}; see section~\ref{S_D_eqs}. This system of equations was also used for an alternative derivation of the autonomous equations, which were introduced for the first time in paper~\cite{we} for a massless scalar field, generalized for a massive scalar field in~\cite{we2,KP2026}, and used for some cosmological calculations in~\cite{Bhat0,Bhat,Bhat1}. We reviewed this technique of autonomous equations in section~\ref{autonom_eqs}. Let us point out that our proposed technique requires knowing only the first few terms of the perturbative expansion to obtain finite time-dependent functions that approximate the time evolution of the correlation functions of the stochastic approach reasonably well. In section~\ref{numerical}, we compared the time evolution of correlation functions obtained by solving the corresponding autonomous equations with those coming from the numerical integration of the Fokker--Planck equation of the stochastic approach. These evolutions appear to be very close. In sections~\ref{Borel-LeRoy} and~\ref{singularities}, we combined the technique of resummation of divergent series based on the Borel--Le Roy transformation with that of autonomous equations. Again, the results look astonishingly good in spite of the heuristic origin of the autonomous equations. However, as we know, many methods, not only in theoretical physics but also in mathematics, were born heuristically and later acquired solid foundations. One may recall the renormalization procedure, the use of generalized functions (distributions), different techniques for summation of divergent series, or the umbral calculus~\cite{umbral}. Thus, it is possible that there are some deeper reasons behind the effectiveness of the equations described above. It would be interesting to generalize the methods used in the present paper. For example, one could study correlation functions on backgrounds that are less symmetric than de Sitter space, e.g., the slow-roll inflation instead of pure de Sitter, and also consider more complicated sets of spectator fields.

\end{document}